\documentclass{optica-article}

\journal{opticajournal} % for journals or Optica Open

\articletype{Research Article}

\usepackage{lineno}
\usepackage{amsmath}
\usepackage{mathtools}
\usepackage{booktabs}
\usepackage{soul}
\usepackage[title]{appendix}

\begin{document}

\title{Universal multiport interferometers for post-selected multi-photon gates}

\author{ Alessio Baldazzi,\authormark{*} %\authormark{1,*} 
and Lorenzo Pavesi,}%\authormark{1}}

\address{
%\authormark{1} 
Department of Physics, University of Trento, Italy}

\email{\authormark{*}alessio.baldazzi@unitn.it} %% email address is required; see note below about the corresponding author designation

% use {asbstract*} to suppress the copyright line. Copyright information will be added in production

\begin{abstract*} 
We show how to use universal multiport interferometers' schemes in order to create photonic post-selected Controlled-Z and Controlled-Controlled-Z gates, which are equivalent, modulo single-qubit gates, to Controlled-NOT and Toffoli gates, respectively.
The new proposed method is based on the following ingredients: identical single photons, Mach-Zehnder interferometer networks, single-photon detectors and post-selection.
In particular, by using dual-rail path encoding together with auxiliary paths and single photons, we improve the success probabilities of such gates. This result further proves the complexity and richness of Reck and Clements schemes beyond the usual notions and practices of Boson Sampling.

\end{abstract*}

%%%%%%%%%%%%%%%%%%%%%%%%%%  body  %%%%%%%%%%%%%%%%%%%%%%%%%%
\section{Introduction}
\label{sec:intro}
Quantum Photonics~\cite{harris_large-scale_2016,qiang_large-scale_2018,o2009photonic,laing2010high} offers a promising platform not only for communication~\cite{brassard2003quantum,gisin2007quantum}, but also for computing~\cite{nielsen_chuang_2010,bennett2000quantum,ladd2010quantum}. 
Photons are the natural choice to transfer information because of low decoherence, high speed and a well-established technology that allows the manufacturing of photonic integrated circuit working at room temperature. Moreover, by exploiting their different degrees of freedom~\cite{o2007optical,tan_resurgence_2019}, it is possible to encode the information either in discrete or in continuous variables with a wide variety of choices.
These variables can be manipulated by linear operations with high fidelity, robustness and flexibility. For example, Mach-Zehnder interferometer (MZI)~\cite{cerf_optical_1998,adami_quantum_1999} or MZI network~\cite{reck_experimental_1994,clements_optimal_2016} can be used to implement any transformations of a generic qudit, i.e. d-dimensional qubit, encoded in the spatial degrees of freedom.
However, the linear manipulation alone cannot execute all the gates needed for universal quantum computing (QC). On the photonic platform, universal gates~\cite{divincenzo_two-bit_1995,barenco_elementary_1995,gottesman_heisenberg_1998}, like Controlled-Z (CZ) and Controlled-NOT (CNOT), require a manipulation where photons cannot be treated independently, or equivalently where they get entangled. 

Remarkably, Knill et al.~\cite{knill_scheme_2001} showed that nonlinear transformations, like Kerr gate~\cite{Kerr_milburn}, are not necessary to achieve all the desired universal gates. Linear optical quantum computing (LOQC)~\cite{cerf_optical_1998,o2007optical,carolan2015universal,rudolph2017optimistic,tan_resurgence_2019,wang2020integrated,Bartlett2020Universal,Dong_2023} can be realized using post-selection and/or heralding. By choosing specific criteria and neglecting the events not satisfying those criteria, an effective nonlinearity~\cite{Scheel} emerges and allows to create multi-photon entangling gates. This scenario implies that some events are discarded, and thus the success probability of these gates is lower than 100$\%$, even if the fidelity of the operation could be one.
This new paradigm for photon entangling gates can be adapted to both approaches to quantum algorithms: measurement-based~\cite{briegel2001persistent,raussendorf2001one} and gate-based~\cite{nielsen_chuang_2010} QC. 

On the photonic platform, the first one is named fusion-based LOQC and it is achieved with entanglement resources coming from parametric nonlinear processes~\cite{helt2012does} together with fusion gates~\cite{browne2005resource,bartolucci_fusion-based_2023}. The parametric processes are probabilistic, but researchers are also exploring deterministic entanglers~\cite{larsen2019deterministic,cogan2023deterministic}. The fusion gates act as parity check operations that sacrifice the initial entanglement to create new correlations and then produce large cluster states as a resource for quantum algorithms~\cite{lindner2009proposal,adcock2019programmable}. There are two types of fusion gates and both have a success probability of 50$\%$. Contrary to type-I, when the type-II fusion gate fails, the operation can be repeated by consuming more pairs of correlated photons to increase the success probability asymptotically to $100\%$~\cite{browne2005resource}. The main drawback of this approach and its scalability relies on the need for many initial correlated photons, which are produced by probabilistic processes or through entangler structures that are difficult to integrate.

The gate-based approach on the photonic platform is characterized by probabilistic gates with lower success probabilities with respect to fusion gates. The first example, given by Knill et al.~\cite{knill_scheme_2001}, is the heralded CZ gate that has $1/16$ success probability. It makes use of the Hong-Ou-Mandel (HOM) effect~\cite{hong_measurement_1987} and two non-linear sign gates. 
The non-linear sign gate~\cite{Ralph_2001} is a conditional sign flip operation that is heralded by the click of one single photon and gives a $\pi$ phase shift only to two-photon states. Since there are two non-linear sign gates, the heralded CZ gate requires two auxiliary single photons and four auxiliary paths in total. %\cite{knill2001note}
A second example uses one Bell state as a resource and beam splitters. It achieves CZ gate with $1/4$ success probability~\cite{pittman2001probabilistic}: such a high value is obtained by increasing the requirement on the initial resources.
On the other hand, Ref.~\cite{postsel_CZ} introduces the most resource-efficient design of a gate-based LOQC CZ gate. The dual-rail structure is modified by the addition of two auxiliary waveguides and, through three beam splitters and post-selection, the desired output has  $1/9$ probability of success. The main limitation of this gate design relies on the fact that it is not a heralded operation and thus it cannot be applied twice to two pairs of qubits that share at least one qubit. In~\cite{Kwon_24}, it is shown how to execute this gate on two pairs of qubits that share only one qubit through the {\it truncation trick}.
In~\cite{bao2007optical,li2021heralded}, it is presented and experimentally demonstrated a controlled-NOT gate, which utilizes two single photons and has $1/8$ success probability. 
The same performance in terms of success probability is found for the scheme reported in~\cite{liu2022universal}. However, in this case there are no auxiliary photons: the result is based on the use of both polarization and spatial degrees of freedom.
Lastly, Ref.~\cite{liu2023linear} shows a post-selected CZ gate with $1/4$ success probability by using one auxiliary single photon and feed-forward operation.

This review of CZ gate designs shows that despite a huge effort there is still the need to achieve high success probability and optimize the required resources. 
The same situation is found for other important universal gates such as the Controlled-Controlled-Z (CCZ) gate and Toffoli gate~\cite{toffoli}. These gates are equivalent modulo single-qubit gates. The decomposition of Toffoli gate requires six two-qubit controlled gates~\cite{shende2008cnotcost}, thus it is essential to find direct ways to implement it in order to have more resource-efficient hardware.
Using linear optics, post-selected Toffoli gate designs have been proposed with success probabilities $1/133$~\cite{fiuravsek2006linear} and $1/72$~\cite{ralph2007efficient} without auxiliary photons.
The use of two Bell states increases the success probability to $1/32$~\cite{ralph2007efficient}. 
Then, like in the case of CZ gate, Ref.~\cite{liu2022universal} presents how to use auxiliary degrees of freedom to obtain $1/64$ success probability.
A higher success probability of $1/60 \approx 1.7\%$~\cite{li2022quantum} can be achieved by using a $6\times 6$ non-unitary MZI network and no initial entanglement resources.
Finally, utilizing two auxiliary single photons and feed-forward operation, Ref.~\cite{liu2023linear} reaches $1/30$ success probability.

In this framework, our work aims to improve the performances of post-selected photonic multi-qubit gates by using universal multiport interferometers, i.e. Reck~\cite{reck_experimental_1994} and Clements~\cite{clements_optimal_2016} schemes. 
Such photonic networks are the fundamental ingredients to perform
% The idea rests on the observation that 
Boson Sampling~\cite{boson_sampling,boson_sampling_review}, which is typically used as an example of quantum analogical simulation able to show quantum advantage for hard computational problems. On one side Gaussian Boson Sampling~\cite{hamilton2017gaussian} can be used to solve vibrational structures of molecular systems, and on the other side scattershot Boson Sampling~\cite{wang2019boson} estimates matrices permanents.
%without any concrete utility. This is the main reason why scattershot Boson Sampling is considered just an academic exercise by many researchers.
%This position has stimulated the community to look for new applications of this procedure, and an interesting example is found in key-distribution~\cite{wang2023experimental}.
Indeed, even if the involved transformations are linear, the dynamics of Boson Sampling is highly non-trivial, since the interference between indistinguishable outputs produces counter-intuitive results. The simplest example is given by HOM effect~\cite{hong_measurement_1987}, where two single photons are inserted in the inputs of a balanced beam splitter producing an entangled path-encoded state~\cite{bouchard2020two,branczyk2017hong}. Therefore, MZI networks are rich of potential interesting transformations even if they have a linear action, and thus they deserve more investigations.
The complexity of MZI networks suggests that multi-photon gates can be obtained by suitable choices of the MZIs' phases and by post-selection criteria, that preserve the chosen qubit structure.
In addition, hybrid quantum-classical computation methods such as the VQE~\cite{peruzzo2014variational,mcclean2016theory} can benefit from improved photonic entangling gates to prepare the trial states. Variational quantum algorithms~\cite{Cerezo_2021} do not require deep circuits or quantum error-correction codes, thus they provide examples where the low success probabilities of our proposed gates could find a practical application.

This paper is organized as follows.
Section~\ref{sec:qubitstruc} introduces the dual-rail path encoding with auxiliary paths and single photons. Section~\ref{sec:cz} and Section~\ref{sec:ccz} show how to implement probabilistic photonic CZ, CNOT, CCZ and Toffoli gates with linear optics and post-selection. Two-qubit gates, CZ and CNOT, have $15.2\%$ success probability and require one auxiliary path and single photon, while three-qubit gates, CCZ and Toffoli show a success probability of $2.7\%$ and two auxiliary paths and single photons.
In Section~\ref{sec:conclu} we compare our results with the literature.

\section{Dual-rail path encoding with auxiliary paths and single photons}
\label{sec:qubitstruc}

\begin{figure}
    \centering
    \includegraphics[width=.5\linewidth]{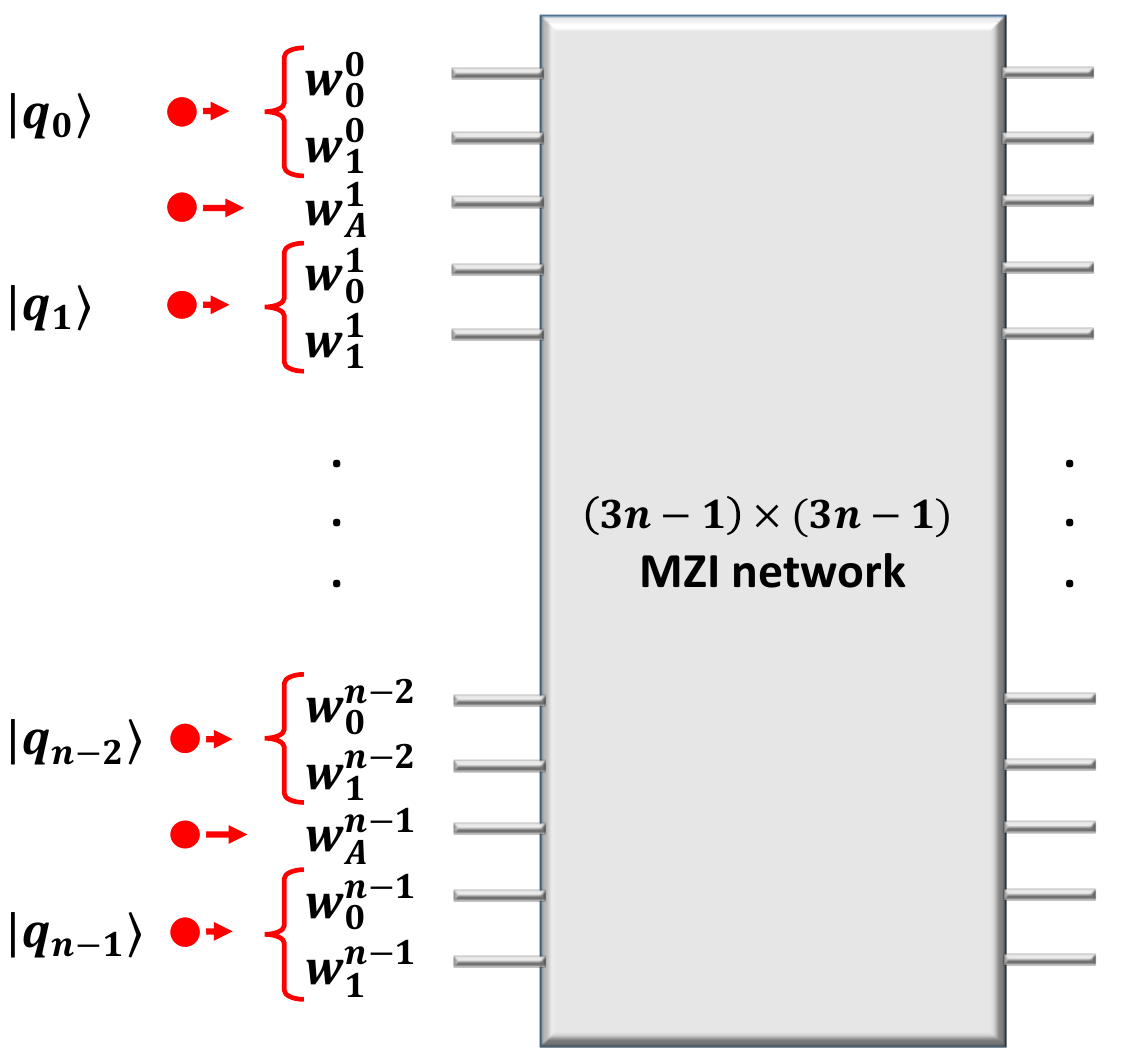}
    \caption{Graphical representation of the new dual-rail qubit structure: auxiliary waveguides, denoted by $\{w^j_A\}_{j\in(1\ldots n-1)}$, are inserted between every pair of path-encoded qubits.
    The red balls represent identical photons and waveguides $(w^j_0,w^j_1)$ are the paths associated respectively to the $j$-th qubit's computational basis states $\vert \mathsf{0} \rangle_j$ and $\vert \mathsf{1} \rangle_j$. 
    The manipulation of $n$ path-encoded qubits is performed by $(3n-1) \times(3n-1)$ MZI network together with the post-selection, i.e. the output events must preserve the chosen qubit structure that is set in input.}
    \label{fig:new_scheme}
\end{figure}

We decide to use path encoding: a physical qubit is represented by the position of one photon in a pair of waveguides, denoted as $(w_{0},w_{1})$.\\
The computational basis states are defined as follows
\begin{equation}
    \begin{aligned}
    \vert \mathsf{0} \rangle &\equiv \hat{a}_{w_{0}}^\dagger \vert \Omega \rangle = \vert \textbf{1}, \textbf{0} \rangle_{\left(w_{0}, w_1\right)} \,,\\
    \vert \mathsf{1} \rangle &\equiv \hat{a}_{w_1}^\dagger \vert \Omega \rangle = \vert \textbf{0}, \textbf{1} \rangle_{\left(w_{0}, w_1\right)} \,,
    \end{aligned}
    \label{eq:qubit_assigned}
\end{equation}
where $a^\dagger$s are creation operators and bold numbers denote occupation numbers and $|\Omega\rangle$ is the vacuum state.
Then, we put an additional waveguide with one auxiliary photon between all adjacent pairs of waveguides assigned to physical qubits.
Therefore, $n$ qubits require $3n-1$ waveguides and $2n-1$ identical photons: $2n$ waveguides and $n$ photons assigned to the path-encoded qubits and $n-1$ auxiliary waveguides and photons.
Figure~\ref{fig:new_scheme} shows the proposed regularly-labeled structure for $n$ qubits together with the manipulation stage, made of a $(3n-1) \times(3n-1)$ MZI network.

The generic path-encoded $n$-qubit state reads
\begin{equation}
    \begin{aligned}
        \vert \Psi \rangle &= \bigotimes_{j=0}^{n-1} \vert \psi \rangle_{j} = \bigotimes_{j=0}^{n-1}  \left( \alpha_{j} \, \vert \mathsf{0} \rangle_j + \beta_{j} \, \vert \mathsf{1} \rangle_j \right) \\
        & \iff \prod_{k=1}^{n-1} \hat{a}^\dagger_{w_A^{k}} \prod_{j=0}^{n-1} \left( \alpha_{j} \, \hat{a}^\dagger_{w_0^{j}} + \beta_{j} \, \hat{a}^\dagger_{w_1^{j}} \right) \vert \Omega \rangle \,,
    \end{aligned}
    \label{eq:generic_nqstate}
\end{equation}
where the index $j$ denotes the $j$-th pair of waveguides which is assigned to the the $j$-th path-encoded qubit and it holds $\vert \alpha_{j} \vert^2 + \vert \beta_{j} \vert^2 = 1$.
The first product refers to the auxiliary photons in the additional waveguides and the second one to the photons assigned to the path-encoded qubits.

The generic manipulation of the dual-rail path-encoded qubit is achieved by considering the MZI as a fundamental building block. The MZI has two inputs and two outputs, thus considering the vector of creation operators of the inputs, $(a^\dagger_{w_0},a^\dagger_{w_1})$, the MZI action can be described with the following matrix 
\begin{equation}
    \begin{split}
    U_{\rm MZI}( t , \phi )  
    &= 
    \begin{pmatrix}
        {\rm e}^{{\rm i}\phi} \, t & \sqrt{1-t^2} \\
        {\rm e}^{{\rm i}\phi}\,\sqrt{1-t^2} & -t
    \end{pmatrix} \,
    \end{split}
\label{eq:MZI_matrix}
\end{equation}
modulo global phases. Here $t\equiv \sin(\theta/2)$, $\theta$ is the phase difference between the two internal MZI arms and $\phi$ is the phase difference imposed between the entries of the MZI. These two phases allow to manipulate the generic dual-rail path-encoded single-qubit state. 
By exploiting the control on external parameters like temperature, pressure, etc., we can create phase differences between the paths assigned to the photonic path-encoded single-qubit state. In this way, it is possible to manipulate and reconfigure single MZIs and MZI networks.
In Appendix~\ref{app:preliminaries}, it is shown how to obtain reconfigurable integrated beam splitters, described by Equation~\ref{eq:MZI_matrix}, with integrated MZIs and how to embed such $2\times 2$ transformation in a generic MZI network with $m$ modes. 
MZI networks have two different designs: Reck~\cite{reck_experimental_1994} and Clements~\cite{clements_optimal_2016} schemes. 
Given a generic set of $m$ waveguides, the universal transformation can be implemented by these $m\times m$ network of MZIs. The term ``universal'' does not mean that such a network can deterministically execute QC universal gates like CZ and CNOT. It means that any input single-photon state propagating in the $m$-modes circuit can be mapped to any output single-photon state. More rigorously, the MZI network implements a generic unitary transformation for $m$-dimensional complex vector, which represents the probability amplitudes of the single photon propagating in $m$ paths. For example, given a single photon inserted in the $j$-th input, the transformation is
\vspace{-0.2cm}
\begin{equation}
    \hat{a}^\dagger_{j}
    \to
    \sum_{k=1}^m \,u^{-1}(j,k) \,\hat{a}^\dagger_{k}
    \implies
    \vert \overbrace{\textbf{0}\ldots\textbf{0}}^{j-1} \textbf{1}   ,\textbf{0}\ldots\textbf{0} \rangle_{\left(w_1\ldots w_m\right)}
    \to 
    \sum_{k=1}^m \,u^{-1}(j,k) \vert \overbrace{\textbf{0}\ldots\textbf{0}}^{k-1} \textbf{1} \,\textbf{0}\ldots\textbf{0}  \rangle_{\left(w_1\ldots w_m\right)}
    \label{eq:genprop}
\end{equation}
where $\hat{a}^\dagger$s are the creation operator of the network modes $\left(w_1\ldots w_m\right)$ and $u^{-1}(j,k)$ are the inverse matrix elements that describe the action of the MZIs network. The previous equation tells us that the input state with a photon in the $j$-th waveguide is transformed into a generic superposition state of the photon in the set of $m$ waveguides and the amplitude probability related to the photon in output $k$ given input $j$ is exactly the component $u^{-1}(i,k)$.
Thus, summarizing our manipulation operative procedure, single-qubit gates are deterministically achieved through a MZI acting on a specific single photon propagating in the corresponding two paths assigned to its path-encoded qubits. Instead, for two- and three-qubit gates we exploit MZI networks acting on two or three path-encoded qubits together with one and two auxiliary photons, respectively.
In the next sections, we show how to implement CZ and CCZ gates that satisfy the presented qubit structure.

After the manipulation post-selection is imposed in the measurement stage. The post-selection criterion consists of considering only output events that satisfied the qubit structure reported in Equation~\eqref{eq:generic_nqstate}. In particular, for $n$-qubit system, the considered events involve the detection of $2n-1$ coincident photons: one photon from each pair of waveguides associated with the path-encoded qubits and one photon from each auxiliary waveguide.
Ideally, the qubit structure is always preserved by single-qubit gates, while the two- and three-qubit gates rely on post-selection. Thus, they are characterized by a success probability lower than 100$\%$ even in the ideal case. In real devices, additional losses of photons produce events that do not satisfy the qubit structure for any path-encoded gate.
These events are discarded and contribute to a further lower success probability. We point out that post-selection measurements can be done through threshold detectors, since only events with a number of coincident clicks equal to the number of single photons present in the systems are considered.

\section{Post-selected CZ and CNOT gates in $5\times 5$ MZI schemes}
\label{sec:cz}

\begin{figure}
    \centering
    \includegraphics[width=0.75\linewidth]{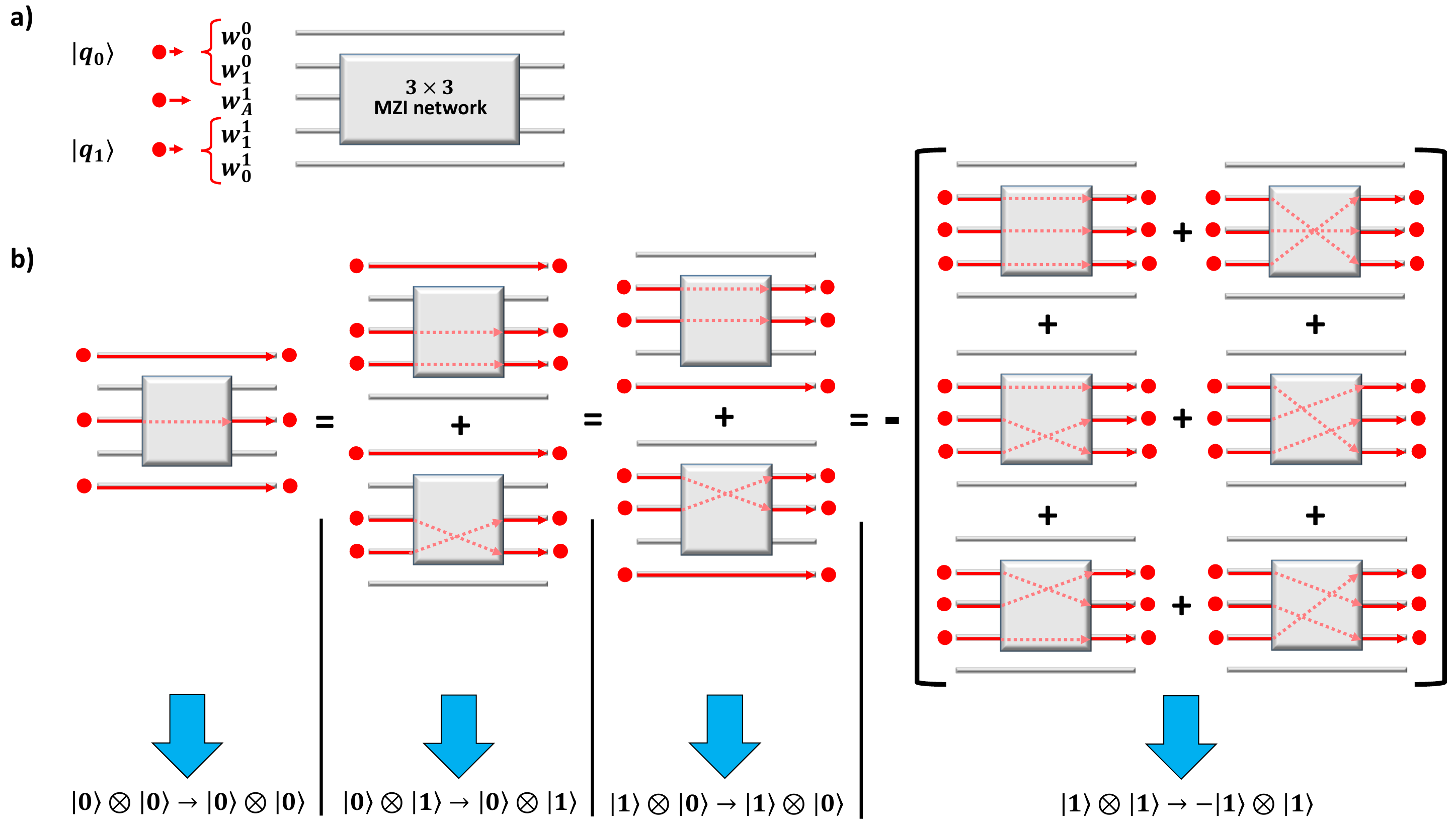}
    \caption{
    (a) Graphical representation of the new dual-rail qubit structure for two qubits: the auxiliary waveguide, denoted by $w_A$, is inserted between every pair of path-encoded qubits.
    The red balls represent identical photons and waveguides $(w^j_0,w^j_1)$ are the paths associated respectively to the $j$-th qubit computational basis states $\vert \mathsf{0} \rangle_j$ and $\vert \mathsf{1} \rangle_j$.
    Note that with respect to Figure~\ref{fig:new_scheme} the waveguides of the second qubit, i.e. $(w^1_0,w^1_1)$, are swapped and the manipulation is simplified by considering a $3\times 3$ MZI network among modes $(w^0_1,w_A,w^1_1)$.
   (b) Graphical representation of desired input-output configurations present in Equation~\eqref{eq:cz_cond}. Such conditions on the probability amplitudes ensure the correct behavior for the post-selected CZ gate, which involves a $3\times 3$ MZI scheme acting only on modes $(w_1^{0}, w_A, w_1^{1})$. In the text, it is shown how to make such structure regularly labeled simply by using swap operations on second qubit paths, $(w_0^{1}, w_1^{1})$.}
    \label{fig:perm_cz}
\end{figure}

In this section, we describe a new design to achieve post-selected CZ and CNOT gates through a $5\times 5$ MZI network. In particular, we present the operative setting of $5\times 5$ Reck and Clements schemes as universal two-qubit gates. Appendix~\ref{app:controlledgates} reports the logic tables of CZ and CNOT gates.

The generic initial state with two qubits is
\begin{equation}
    \vert \Psi \rangle_{2,\rm in} = \vert q_0 \rangle_{\rm in} \otimes \vert q_1 \rangle_{\rm in} \,, 
    \quad \mbox{where}\quad
    \begin{cases}
        \vert q_0 \rangle_{\rm in} &=
     \alpha_{0} \, \vert \mathsf{0} \rangle_0 + \beta_{0} \, \vert \mathsf{1} \rangle_0 \,,\\
     \vert q_1 \rangle_{\rm in} &= 
     \alpha_{1} \, \vert \mathsf{0} \rangle_1 + \beta_{1} \, \vert \mathsf{1} \rangle_1 \,.
    \end{cases}
\label{eq:generic_2qstate}
\end{equation}
% where $\alpha^2 +\beta^2 = 1$.\\
Following the assignment described in the previous section, the initial path-encoded state reads
\begin{equation}
    \vert \Psi \rangle_{2,\rm in}
    \iff
    \left( \alpha_{0} \,\hat{a}_{w_{0}^{0}}^\dagger + \beta_{0} \,\hat{a}_{w_{1}^{0}}^\dagger \right)
    \,\hat{a}_{w_{A}}^\dagger \,
    \left( \alpha_{1} \,\hat{a}_{w_{0}^{1}}^\dagger + \beta_{1} \,\hat{a}_{w_{1}^{1}}^\dagger \right) \vert \Omega \rangle\,.
\end{equation}

The strategy to find the new post-selected CZ gate consists of exploiting the interference between the auxiliary photon and the photons in waveguides assigned to the states $|\mathsf{1} \rangle$.
In particular, by utilizing the following non-regularly-labeled order 
\begin{equation}
    (\hat{a}^\dagger_{w_0^{0}}, \hat{a}^\dagger_{w_1^{0}}, \hat{a}^\dagger_{w_A}, \hat{a}^\dagger_{w_1^{1}}, \hat{a}^\dagger_{w_0^{1}}) \,,
\end{equation}
we can focus on $3\times 3$ MZI schemes, which act on modes $(w_1^{0}, w_A, w_1^{1})$. Figure~\ref{fig:perm_cz}(a) shows a graphical representation of the non-regularly-labeled qubit structure and the used $3\times 3$ MZI scheme.
Note that if we denote the matrix associated with the $3\times 3$ MZI scheme by $u_3$, the component $u_3^{-1}(j,m)$ is the probability amplitude of the transition of one photon from mode $j$ to mode $m$ of the $3$ waveguides given by modes $(w_1^{0}, w_A, w_1^{1})$.

By decomposing the $3\times 3$ MZI scheme associated with $u_3$ in the $2\times 2$ MZI matrix described by Equation~\eqref{eq:MZI_matrix}, we have a reconfigurable network where the desired input-output configuration can be reached by a proper setting of the MZIs phases.
To implement the CZ gate, the unitary matrix $u_3$ must satisfy the following four conditions
\begin{equation}
    u_3^{-1}(2,2) = \mbox{perm}\left[ u_3^{-1}(1,2;1,2) \right] 
    = \mbox{perm}\left[ u_3^{-1}(2,3;2,3) \right]
    = -\mbox{perm}\left[ u_3^{-1} \right] \,,
    \label{eq:cz_cond}
\end{equation}
where 
%$u(i,m)$ is the component of $u$ in position $(i;m)$ and 
$u(i_1,\ldots,i_j;m_1,\ldots,m_n)$ indicates the square sub-matrix of $u$ with rows $[i_1,\ldots,i_j]$ and columns $[m_1,\ldots,m_n]$.
Equation~\eqref{eq:cz_cond} are graphically represented in Figure~\ref{fig:perm_cz}.
Note that we consider only inputs and outputs that preserve the chosen qubit structure and we impose a minus sign only when the three photons enter the $3\times 3$ MZI scheme, or equivalently only when both input qubits are in the state $|\mathsf{1} \rangle$. This is equivalent to obtain the CZ gate on the restricted set of events that preserve the qubit structure.
%Equation~\eqref{eq:generic_nqstate}(b).

Equation~\eqref{eq:cz_cond} describe a $3 \times 3$ matrix whose permanent is opposite to the permanent of all its sub-matrices. Thus, these conditions together with unitarity are mathematically non-trivial.

The generic manipulation of the three central modes, as reported in Figure~\ref{fig:perm_cz}, can be decomposed into 3 MZIs as follows
($3\times 3$ Reck and Clements schemes have identical structure)
\begin{equation}
    \bar{U}_5(\{ t_k , \phi_k \}_{k=1\ldots 3}) \equiv U_5^{(2,3)}(t_3, \phi_3) \cdot U_5^{(3,4)}(t_2, \phi_2) \cdot U_5^{(2,3)}(t_1, \phi_1) \,,
\end{equation}
where we are using Equation~\eqref{eq:MZI_matrix} and the subscript denotes the total number of modes in the network and the superscript the pair of modes connected by the MZI.
Then, 
% we simplify the structure by putting to zero all $\phi$s and 
we add the swapping transformation $U_5^{(4,5)}(0, 0)$ to have the regularly-labeled qubit structure reported in Figure~\ref{fig:new_scheme}, and we obtain the $5\times 5$ network represented by the following matrix
\begin{equation}
    U_{5}^{\rm }%(\{ t_k , \phi_k \}_{k=1\ldots 3}) 
    \equiv
    X_5^{(4,5)} \cdot \bar{U}_5(\{ t_k , \phi_k \}_{k=1\ldots 3}) \cdot X_5^{(4,5)} \,,
\end{equation}
where $X_m^{(i,j)}\equiv U_m^{(i,j)}(0, 0)$ is the swap operation between modes $(i,j)$ in $m$-mode network. 
% Note that $U_5^{(45)}(0, 0)$ swaps the position of $\hat{a}^\dagger_{w_0^{1}}$ $\hat{a}^\dagger_{w_1^{1}}$.

% Then, we construct networks of MZIs by using the recipe of Reck and Clements. In the case of two qubits, we need $5\times 5$ MZI networks, represented in Figure~\ref{fig:5x5}.

% \begin{figure}
%     \centering
%     \includegraphics[width=\linewidth]{reck_clements.pdf}
%     \caption{a) Representation of a single MZI: the blue parts are waveguides and integrated beam splitter, while the red parts are phase shifters. b) The new path encoding structure for a two-qubit system embedded in the $5\times 5$ Clements scheme. c) The new path encoding structure for a two-qubit system embedded in the $5\times 5$ Reck scheme.}
%     \label{fig:5x5}
% \end{figure}

Applying the conditions in Equation~\eqref{eq:cz_cond} to the central $3\times 3$ submatrix of $\bar U_{5}^{\rm }$, it is possible to find the following solution:
\begin{equation}
    (t^*_1,t^*_2,t^*_3) \approx ( 0.3686, -0.2192, 0.8686 ) \,,
    \label{eq:special_setting}
\end{equation}
and all the $\phi$s equal to zero. This solution is unique modulo $t_1^* \leftrightarrow t_3^*$ and $t_{1/3}^* \to - t_{1/3}^*$.
As it is explained in Appendix~\ref{app:preliminaries}, negative transmittance values mean that the photon is transmitted with a ratio given by the squared value of the transmittance and the reflected and transmitted amplitudes of the photon state have an additional $\pi$ shift difference.

\begin{figure}
    \centering
    \includegraphics[width=\linewidth]{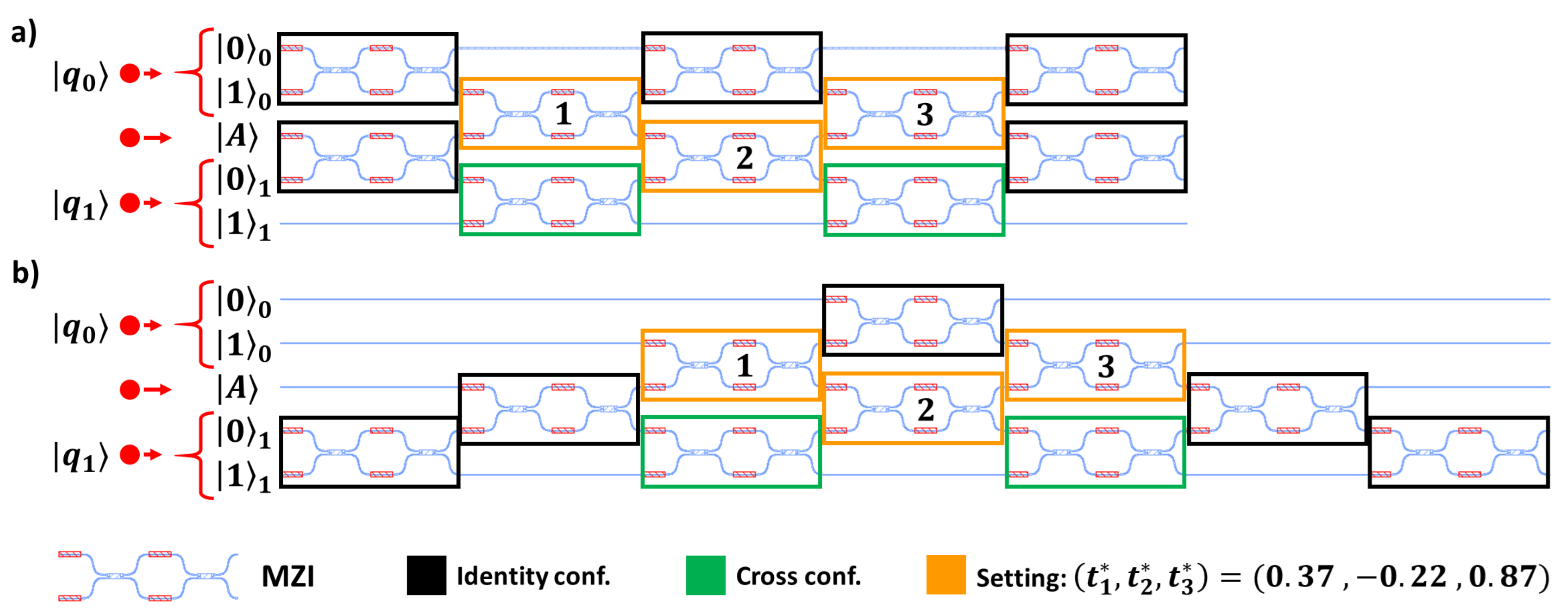}
    \caption{ The $5\times 5$ MZI networks with (a) Clements and (b) Reck schemes, considered to implement the new post-selected CZ using the qubit structure shown in Figure~\ref{fig:new_scheme}.
    The red balls represent identical photons, $|A\rangle$ is the state of one auxiliary photon in the auxiliary waveguide inserted between the path-encoded qubits.
    The networks are composed of MZIs: light blue parts are waveguides and red parts are phase shifters. MZI unit acts as a tunable beam splitter, described by Equation~\eqref{eq:MZI_matrix}.
    The colors of the borders stand for the set configuration of each MZI: black means identity, green means cross or swap and orange stands for the setting reported in Equation~\eqref{eq:special_setting}, where $t$-parameter is the MZI transmittance. 
    The numbers $t_j$ refer to the transmittance implemented by the MZI labeled with $j$.
    The negative transmittance values mean that the photon is transmitted with a ratio given by the squared value of the transmittance and the reflected and transmitted amplitudes of the photon state have an additional $\pi$ shift difference.
    }
    \label{fig:CZnew}
\end{figure}

Therefore, by using the preliminary notions given in Appendix~\ref{app:preliminaries}, the state evolution associated with the new post-selected CZ can be written as
\begin{equation}
    {\rm V}_2
    \to
    \left( U_{5}^{\rm CZ} \right)^{-1}
    \cdot
    {\rm V}_2 \,,
    \label{eq:recipeCZ}
\end{equation}
where
\begin{equation}
\begin{split}
    {\rm V}_2^{\rm T} &\equiv 
    \left(
        \hat{a}_{w_{0}^{0}}^\dagger,
        \hat{a}_{w_{1}^{0}}^\dagger,
        \hat{a}_{w_{A}^{}}^\dagger,
        \hat{a}_{w_{0}^{1}}^\dagger,
        \hat{a}_{w_{1}^{1}}^\dagger
    \right) \,,\\
U_{5}^{\rm CZ} &\equiv 
% \footnotesize
\left(
\begin{array}{ccccc}
 1 & 0 & 0 & 0 & 0 \\
 0 & 0.2192 & 0.8475 & 0 & 0.4834 \\
 0 & 0.3597 & 0.3904 & 0 & -0.8475 \\
 0 & 0 & 0 & 1 & 0 \\
 0 & 0.9070 & -0.3597 & 0 & 0.2192 \\
\end{array}
\right) \,,
\end{split}
\end{equation}
and the superscript ``T'' denotes the transpose.
In Figure~\ref{fig:CZnew}(a-b) we show such MZI network embedded in $5\times 5$ Clements and Reck schemes, respectively.
% 'Boson Sampling is thus used to compute the permanents in Equation\eqref{eq:genprop} by using the qubits evolution~\eqref{eq:recipeCZ}.'

Applying the linear transformation in Equation~\eqref{eq:recipeCZ}, the state $\vert \Psi \rangle_{2,\rm in}$ evolves to $\vert \Psi \rangle_{2,\rm out}$, which reads modulo global phase as
\begin{equation}
    \begin{aligned}
        \vert \Psi \rangle_{2,\rm out}
        =&\, {\rm A}^{\rm CZ}_{\rm succ} \left(   \alpha_{0} \alpha_{1} \,\hat{a}_{w_{0}^{0}}^\dagger \hat{a}_{w_{0}^{1}}^\dagger + \alpha_{0}\beta_{1}\,\hat{a}_{w_{0}^{0}}^\dagger\hat{a}_{w_{1}^{1}}^\dagger 
        + \beta_{0} \alpha_{1}\,\hat{a}_{w_{1}^{0}}^\dagger \hat{a}_{w_{0}^{1}}^\dagger - \beta_{0} \beta_{1} \,\hat{a}_{w_{1}^{0}}^\dagger \hat{a}_{w_{1}^{1}}^\dagger\right) \,\hat{a}_{w_{A}^{}}^\dagger \,\vert \Omega \rangle 
        +\ldots
    \end{aligned}
    \label{eq:cz_hp}
\end{equation}
where ${\rm A}^{\rm CZ}_{\rm succ} \approx 0.3904$ and the dots contain all the terms that do not preserve the initial qubit structure. 
% In particular, among these terms we find events with 
% \begin{enumerate}
%     \item one photon in $w_{A}^{0}$ and one photon in the doublet $( w_{0}^{0}, w_{1}^{0})$ or $( w_{0}^{1}, w_{1}^{1})$,
%     \item one photon in $w_{A}^{1}$ and one photon in the doublet $( w_{0}^{0}, w_{1}^{0})$ or $( w_{0}^{1}, w_{1}^{1})$,
%     \item two photons in the auxiliary waveguides $( w_{A}^{0}, w_{A}^{1})$,
%     \item two photons in the doublet $( w_{0}^{0}, w_{1}^{0})$ or $( w_{0}^{1}, w_{1}^{1})$.
% \end{enumerate}
% The post-selection discards these events. In the presence of losses, we also discard events with one or no photon in the six waveguides set.

Finally, we can rewrite Equation~\eqref{eq:cz_hp} using the assignment with computational basis%, given Equation~\eqref{eq:basis}:
\begin{equation}
    \begin{aligned}
        \vert \Psi \rangle_{2,\rm out} \propto \alpha_{0} \alpha_{1} \vert \mathsf{00} \rangle + \alpha_{0} \beta_{1} \vert \mathsf{01} \rangle + \beta_{0} \alpha_{1} \vert \mathsf{10} \rangle - \beta_{0} \beta_{1} \vert \mathsf{11} \rangle  \,,
    \end{aligned}
    \label{eq:cz_corr}
\end{equation}
where we neglect the contributions that don't preserve the qubit structure and are not considered after the post-selection.

The success probability is 
\begin{equation}
    \mathbb{P}_{\rm succ}^{\rm CZ} = \,\Big|{\rm A}^{\rm CZ}_{\rm succ}\Big|^2 \approx 15.2 \%\,.
    \label{eq:cz_succ}
\end{equation}

By taking $| q_0\rangle$ as the control qubit and $| q_1\rangle$ as the target qubit, CNOT gate can be simply achieved by adding to the previous scheme single-qubit gates as follows
\begin{equation}
    U_{5}^{\rm CNOT} = 
    U_{12}(1, 0) \cdot U_{45}(1/\sqrt{2}, 0) \cdot \bar{U}_5(\{ t_k^* , \phi_k^* \}_{k=1\ldots 3})  \cdot U_{45}(1/\sqrt{2}, 0) \,,
    \label{eq:cnot_ps}
\end{equation}
where $t^*_k$ are reported in Equation~\eqref{eq:special_setting} and $\phi^*$ are zero.
Note that such operation can be embedded into $5\times 5$ Clements scheme: looking at Figure~\ref{fig:CZnew}, the MZIs inside green boxes must be set in the phase setting $(1/\sqrt{2}, 0)$ and the upper-right MZI with setting $(1,0)$. The $5\times 5$ Reck scheme does not allow such simple embedding, but only the single-qubit gate $U_{12}(1, 0)$ would be missing.\\
It is possible to verify that the previous transformation produces the following output state
\begin{equation}
    \begin{aligned}
        \vert \Psi \rangle_{2,\rm out}
        =&\, {\rm A}^{\rm CZ}_{\rm succ} \left(   \alpha_{0} \alpha_{1} \,\hat{a}_{w_{0}^{0}}^\dagger \hat{a}_{w_{0}^{1}}^\dagger + \alpha_{0}\beta_{1}\,\hat{a}_{w_{0}^{0}}^\dagger\hat{a}_{w_{1}^{1}}^\dagger 
        + \beta_{0} \alpha_{1}\,\hat{a}_{w_{1}^{0}}^\dagger \hat{a}_{w_{1}^{1}}^\dagger + \beta_{0} \beta_{1} \,\hat{a}_{w_{1}^{0}}^\dagger \hat{a}_{w_{0}^{1}}^\dagger\right) \,\hat{a}_{w_{A}^{}}^\dagger \,\vert \Omega \rangle 
        +\ldots
    \end{aligned}
    \label{eq:cnot_hp}
\end{equation}
where the dots contain all the terms that do not preserve the initial qubit structure. 
The previous state is assigned to the correct output state of the CNOT gate,
\begin{equation}
    \begin{aligned}
        \vert \Psi \rangle_{2,\rm out} \propto \alpha_{0} \alpha_{1} \vert \mathsf{00} \rangle + \alpha_{0} \beta_{1} \vert \mathsf{01} \rangle + \beta_{0} \alpha_{1} \vert \mathsf{11} \rangle + \beta_{0} \beta_{1} \vert \mathsf{10} \rangle  \,,
    \end{aligned}
    \label{eq:cnot_corr}
\end{equation}
when we consider only the output configurations preserving the qubit structure.

We conclude this section with two observations.
First, such post-selected CZ gate cannot be cascaded on qubits' pairs that share both qubits. However, it is possible to cascade this operation on qubits' pairs that share only one qubit without using the {\it truncation trick}, explained in \cite{Kwon_24} for the post-selected CZ gate with $1/9$ success probability~\cite{postsel_CZ}. This improvement is made possible by the presence of the auxiliary photon and it is shown in Appendix~\ref{app:cascade}.
Second, the same MZI scheme can be tuned with different settings to implement a post-selected CZ gate with more than one auxiliary photon in the auxiliary waveguide. Indeed, besides the solution for one auxiliary photon given in Equation~\eqref{eq:special_setting}, there are other solutions as we increase the number of auxiliary photons. These are reported in Appendix~\ref{app:auxtower}. Adding such resources implies the use of number-resolving detectors, since the qubit structure becomes characterized by more single photons in each auxiliary waveguide. However, such solutions are characterized by a decreasing success probability as the resources increase. Thus, there is no advantage in having more than one auxiliary photon in each auxiliary waveguide.

\section{Post-selected CCZ and Toffoli gates in $8\times 8$ MZI schemes}
\label{sec:ccz}

\begin{figure}
    \centering
    \includegraphics[width=\linewidth]{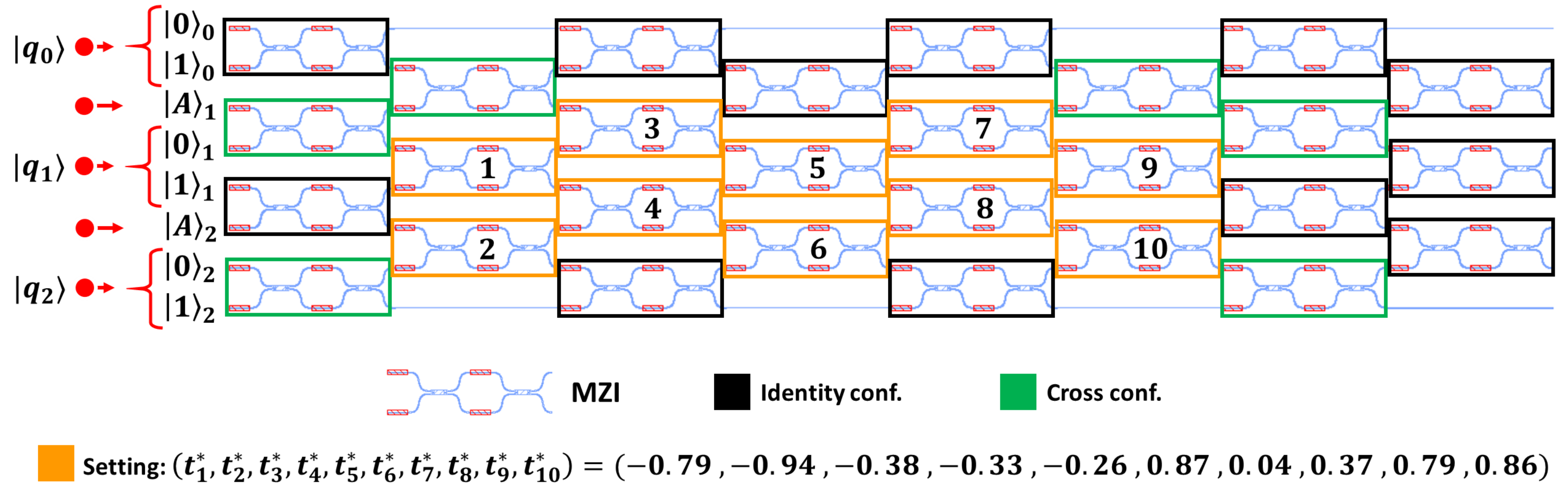}
    \caption{The $8\times 8$ MZI network with Clements scheme, considered to implement the new post-selected CCZ  using the qubit structure shown in Figure~\ref{fig:new_scheme}.
    The red balls represent identical photons, $|A\rangle_j$ are the states of one auxiliary photon in the auxiliary waveguide inserted between every path-encoded qubits.
    The network is composed of MZIs: light blue parts are waveguides and red parts are phase shifters. MZI unit acts as a tunable beam splitter, described by Equation~\eqref{eq:MZI_matrix}.
    The colors of the borders stand for the set configuration of the MZIs: black means identity, green means cross or swap and orange stands for the setting reported in Equation~\eqref{eq:special_settingbis}, where $t$ parameter is the MZI transmittance. 
    The numbers $t_j$ refer to the transmittance implemented by the MZI labeled with $j$.
    The negative transmittance values mean that the photon is transmitted with a ratio given by the squared value of the transmittance and the reflected and transmitted amplitudes of the photon state have an additional $\pi$ shift difference.
    }
    \label{fig:CCZnew}
\end{figure}

In this section, we show a new way to achieve post-selected CCZ and Toffoli gates through an integrated $8\times 8$ MZI network. In particular, the operative setting of MZIs in the $8\times 8$ Reck and Clements schemes as universal three-qubit gates is summarized.
As pointed out in Section~\ref{sec:intro}, the decomposition of Toffoli gate acting on three qubits requires six two-qubit gates. However, in our case we have two important issues: the probabilistic nature of the post-selected two-qubit gates and the impossibility to cascade them on the same pair of qubits twice. 
Therefore, the direct implementation of universal three-qubit gates avoids these limitations.
In Appendix~\ref{app:controlledgates}, the logic tables of CCZ and Toffoli gates are presented.

The generic initial state with three qubits is
\begin{equation}
    \vert \Psi \rangle_{3,\rm in} = \vert q_0 \rangle_{\rm in} \otimes \vert q_1 \rangle_{\rm in} \otimes \vert q_2 \rangle_{\rm in} \,, 
    \quad \mbox{where}\quad
    \begin{cases}
        \vert q_0 \rangle_{\rm in} &=
     \alpha_{0} \, \vert \mathsf{0} \rangle_0 + \beta_{0} \, \vert \mathsf{1} \rangle_0 \,,\\
     \vert q_1 \rangle_{\rm in} &= 
     \alpha_{1} \, \vert \mathsf{0} \rangle_1 + \beta_{1} \, \vert \mathsf{1} \rangle_1 \,,\\
     \vert q_2 \rangle_{\rm in} &= 
     \alpha_{2} \, \vert \mathsf{0} \rangle_2 + \beta_{2} \, \vert \mathsf{1} \rangle_2 \,.
    \end{cases}
\label{eq:generic_3qstate}
\end{equation}
% where $\alpha^2 +\beta^2 = 1$.\\
Following the chosen assignment, the initial path-encoded state reads
\begin{equation}
    \vert \Psi \rangle_{3,\rm in}
    \iff
    \left( \alpha_{0} \,\hat{a}_{w_{0}^{0}}^\dagger + \beta_{0} \,\hat{a}_{w_{1}^{0}}^\dagger \right)
    \,\hat{a}_{w_{A}^1}^\dagger \,
    \left( \alpha_{1} \,\hat{a}_{w_{0}^{1}}^\dagger + \beta_{1} \,\hat{a}_{w_{1}^{1}}^\dagger \right) 
    \,\hat{a}_{w_{A}^2}^\dagger \,
    \left( \alpha_{2} \,\hat{a}_{w_{0}^{2}}^\dagger + \beta_{2} \,\hat{a}_{w_{1}^{2}}^\dagger \right)
    \vert \Omega \rangle\,.
\end{equation}

The strategy to find the new post-selected CCZ gate is again based on the interference between the auxiliary photons and the photons in waveguides assigned to the states $|\mathsf{1} \rangle$.
In particular, by starting with the following non-regularly-labeled order 
\begin{equation}
    (\hat{a}^\dagger_{w_0^{0}}, \hat{a}^\dagger_{w_0^{1}}, \hat{a}^\dagger_{w_1^{0}}, \hat{a}^\dagger_{w_A^1}, \hat{a}^\dagger_{w_1^{1}}, \hat{a}^\dagger_{w_A^2}, \hat{a}^\dagger_{w_1^{2}}, \hat{a}^\dagger_{w_0^{2}}) \,,
\end{equation}
we can focus on $5\times 5$ MZI schemes, which act on modes $(\hat{a}^\dagger_{w_1^{0}}, \hat{a}^\dagger_{w_A^1}, \hat{a}^\dagger_{w_1^{1}}, \hat{a}^\dagger_{w_A^2}, \hat{a}^\dagger_{w_1^{2}})$, and whose associated unitary matrix $u_5$ satisfies the following eight conditions
\begin{equation}
\begin{split}
    & \mbox{perm}\left[u_5^{-1}(2,4;2,4)\right] = \mbox{perm}\left[ u_5^{-1}(2,4,5;2,4,5) \right] 
    = \mbox{perm}\left[ u_5^{-1}(2,3,4;2,3,4) \right] =\\
    & = \mbox{perm}\left[ u_5^{-1}(2,3,4,5;2,3,4,5) \right] 
    = \mbox{perm}\left[ u_5^{-1}(1,2,4;1,2,4) \right] =\\
    & = \mbox{perm}\left[ u_5^{-1}(1,2,4,5;1,2,4,5) \right]
    = \mbox{perm}\left[ u_5^{-1}(1,2,3,4;1,2,3,4) \right]
    = -\mbox{perm}\left[ u_5^{-1} \right] \,,
    \end{split}
    \label{eq:ccz_cond}
\end{equation}
where the notation is the same as Equation~\eqref{eq:cz_cond}.
The previous equations can also be graphically represented analogously to Equation~\eqref{eq:cz_cond} with Figure~\ref{fig:perm_cz}. In this case, the number of equations is eight, i.e. the number of input-output configurations for three-qubit gates. Moreover, each term contains the sum of all probability amplitudes corresponding to input-output events allowed by the chosen qubit structure. 
Reading Equation~\eqref{eq:ccz_cond} from top-left to bottom-right, the numbers of contributions contained in each term are 2, 6, 6, 24, 24, 6, 24, 24 and 120, respectively. Note that these values are the factorials of the number of modes of the $5\times 5$ network involved in the specific input-output configuration.
Note also that among the configurations of inputs and outputs that preserve the chosen qubit structure we impose a minus sign only when all the five photons present in the dynamics enter the $5\times 5$ MZI scheme, or equivalently only when all three qubits are in waveguides assigned to the state $|\mathsf{1} \rangle$. In this way, the CCZ gate is obtained on the restricted set of events that preserve the qubit structure, Equation~\eqref{eq:generic_nqstate}.

\begin{figure}
    \centering
    \includegraphics[width=\linewidth]{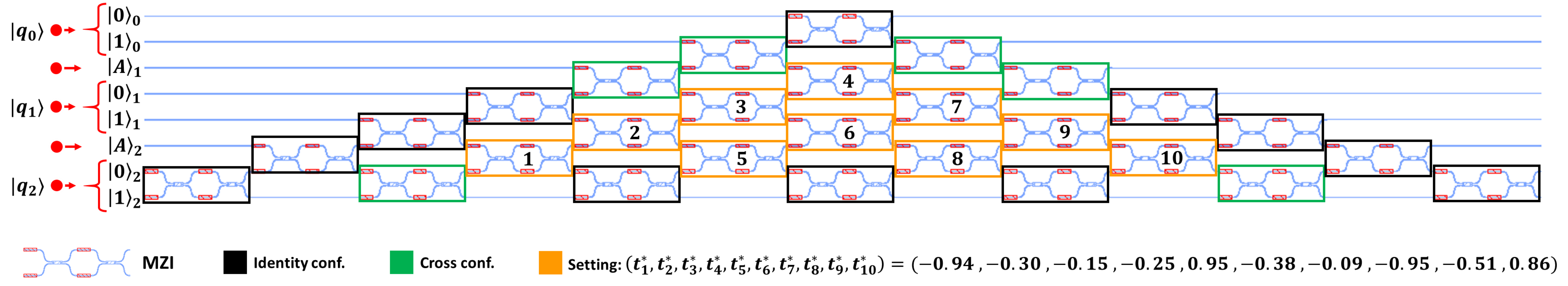}
    \caption{The $8\times 8$ MZI network with Reck scheme, considered to implement the new post-selected CCZ  using the qubit structure shown in Figure~\ref{fig:new_scheme}.
    The red balls represent identical photons, $|A\rangle_j$ are the states of one auxiliary photon in the auxiliary waveguide inserted between every path-encoded qubits.
    The network is composed of MZIs: light blue parts are waveguides and red parts are phase shifters. MZI unit acts as a tunable beam splitter, described by Equation~\eqref{eq:MZI_matrix}.
    The colors of the borders stand for the set configuration of the MZIs: black means identity, green means cross or swap and orange stands for the setting reported in Equation~\eqref{eq:special_settingbis_R}, where $t$ parameter is the MZI transmittance. 
    The numbers $t_j$ refer to the transmittance implemented by the MZI labeled with $j$.
    The negative transmittance values mean that the photon is transmitted with a ratio given by the squared value of the transmittance and the reflected and transmitted amplitudes of the photon state have an additional $\pi$ shift difference.
    }
    \label{fig:CCZnew_R}
\end{figure}

The generic manipulation of the five central modes can be decomposed into 10 MZIs, which are differently set in the two schemes. The Clements scheme is described by the following matrix
\begin{equation}
\begin{split}
    & \bar{U}_{8,c} = U_8^{(4|7)}(t_9, \phi_9, t_{10}, \phi_{10}) \cdot U_8^{\{(4|7),(3|6)\}}(\{ t_k , \phi_k \}_{k=5\ldots 8}) \cdot U_8^{\{(4|7),(3|6)\}}(\{ t_k , \phi_k \}_{k=1\ldots 4}) \\
    &  \mbox{where}\quad  
    \begin{cases}
        U_8^{(4|7)}(t_1,\phi_1,t_2,\phi_2) \equiv U_{8}^{(6,7)}(t_2, \phi_2) \cdot U_{8}^{(4,5)}(t_1, \phi_1) \,,\\
        U_8^{(3|6)}(t_1,\phi_1,t_2,\phi_2) \equiv U_{8}^{(5,6)}(t_2, \phi_2) \cdot U_{8}^{(3,4)}(t_1, \phi_1) \,,\\
        U_8^{\{(4|7),(3|6)\}}(\{ t_k , \phi_k \}_{k=1\ldots 4}) \equiv U_8^{(3|6)}(t_1,\phi_1,t_2,\phi_2) \cdot U_8^{(4|7)}(t_3,\phi_3,t_4,\phi_4) \,,
    \end{cases}
    \label{eq:clem8}
\end{split}
\end{equation}
while for the Reck scheme 
\begin{equation}
\begin{split}
    & \bar{U}_{8,r} = U_8^{(6,7)}(t_{10}, \phi_{10}) \cdot U_8^{(5/7)}(\{ t_k , \phi_k \}_{k=8\ldots 9}) \cdot U_8^{(4/7)}(\{ t_k , \phi_k \}_{k=5\ldots 7}) \cdot U_8^{(3/7)}(\{ t_k , \phi_k \}_{k=1\ldots 4}) \\
    &  \mbox{where}\quad  
    \begin{cases}
        U_8^{(3/7)}(t_1,\phi_1,t_2,\phi_2) \equiv  U_{8}^{(3,4)}(t_4, \phi_4) \cdot U_{8}^{(4,5)}(t_3, \phi_3) \cdot U_{8}^{(5,6)}(t_2, \phi_2) \cdot U_{8}^{(6,7)}(t_1, \phi_1) \,,\\
        U_8^{(4/7)}(t_1,\phi_1,t_2,\phi_2) \equiv  U_{8}^{(4,5)}(t_3, \phi_3) \cdot U_{8}^{(5,6)}(t_2, \phi_2) \cdot U_{8}^{(6,7)}(t_1, \phi_1) \,,\\
        U_8^{(5/7)}(t_1,\phi_1,t_2,\phi_2) \equiv 
        U_{8}^{(5,6)}(t_2, \phi_2) \cdot U_{8}^{(6,7)}(t_1, \phi_1) \,,
    \end{cases}
    \label{eq:reck8}
\end{split}
\end{equation}
where the subscript denotes the total number of modes in the network and the superscript $(i,j)$ the pair of modes $(i,j)$ connected by the MZI.

Then, we add the swapping transformations to have the regularly-labeled qubit structure reported in Figure~\ref{fig:new_scheme}, and we obtain the $8\times 8$ network represented by the following matrix
\begin{equation}
    U_{8,c/r}(\{ t_k , \phi_k \}_{k=1\ldots 10}) = 
    X_8^{(3,4)} \cdot X_8^{(2,3)}\cdot X_8^{(7,8)} \cdot \bar{U}_{8,c/r}(\{ t_k , \phi_k \}_{k=1\ldots 10}) \cdot X_8^{(7,8)} \cdot X_8^{(2,3)}\cdot X_8^{(3,4)} \,,
\end{equation}
where $X_m^{(i,j)}$ is the swap operation between modes $(i,j)$ in $m$-mode network, as above. 

By applying the conditions in Equation~\eqref{eq:ccz_cond} to $U_{8}^{\rm }$ and for simplicity by focusing on solutions with all $\phi$s equal to zero, we find the following solution for the Clements scheme
\begin{equation}
    \begin{aligned}
        & (t^*_1,t^*_2,t^*_3,t^*_4,t^*_5,t^*_6,t^*_7,t^*_8,t^*_9,t^*_{10}) \\
    &= ( -0.7893, -0.9428, -0.3809, -0.3284, -0.2583, 0.8719, 0.03792, 0.3689, 0.7943, 0.8559 ) \,,
    \end{aligned}
    \label{eq:special_settingbis}
\end{equation}
and the corresponding one for the Reck scheme
\begin{equation}
    \begin{aligned}
        & (t^*_1,t^*_2,t^*_3,t^*_4,t^*_5,t^*_6,t^*_7,t^*_8,t^*_9,t^*_{10}) \\
    &= ( -0.9428, -0.3022, -0.1496, -0.2531, 0.9540, -0.3768, -0.09816, -0.9507, -0.5064, 0.8559 ) \,.
    \end{aligned}
    \label{eq:special_settingbis_R}
\end{equation}
Both solutions describe the same unitary transformation.
There are many other non-equivalent solutions to Equation~\eqref{eq:ccz_cond}.
Among those found, we report the ones that have the highest success probability. Moreover, since we consider only real solutions, there is still room for improvement regarding the success probability.

% Contrary to the case of CZ gate explained in Section~\ref{sec:cz}, this solution is not unique, but it is the real solution with higher success probability as far as we know. 

Therefore, the state evolution associated with the new post-selected CCZ can be written as
\begin{equation}
    {\rm V}_3
    \to
    \left( U_{8}^{\rm CCZ} \right)^{-1}
    \cdot
    {\rm V}_3 \,,
    \label{eq:recipeCCZ}
\end{equation}
where
\begin{equation}
\begin{split}
    {\rm V}_3^{\rm T} &\equiv 
    \left(
        \hat{a}_{w_{0}^{0}}^\dagger,
        \hat{a}_{w_{1}^{0}}^\dagger,
        \hat{a}_{w_{A}^{1}}^\dagger,
        \hat{a}_{w_{0}^{1}}^\dagger,
        \hat{a}_{w_{1}^{1}}^\dagger,
        \hat{a}_{w_{A}^{2}}^\dagger,
        \hat{a}_{w_{0}^{2}}^\dagger,
        \hat{a}_{w_{1}^{2}}^\dagger
    \right) \,,\\
U_{8,c/r}^{\rm CCZ} &\equiv 
U_{8,c/r}(\{ t_k^* , 0 \}_{k=1\ldots 10})
% \footnotesize
% \left(
% \begin{array}{cccccccc}
%  1 & 0 & 0 & 0 & 0 & 0 & 0 & 0 \\
%  0 & -0.253097 & -0.144686 & 0 & -0.289053 & -0.859702 & 0 & 0.303922 \\
%  0 & -0.0949594 & -0.367007 & 0 & 0.862646 & -0.0859038 & 0 & 0.323652 \\
%  0 & 0 & 0 & 1 & 0 & 0 & 0 & 0 \\
%  0 & -0.487524 & -0.713273 & 0 & -0.174711 & 0.16601 & 0 & -0.44213 \\
%  0 & 0.710573 & -0.285319 & 0 & 0.110023 & -0.377978 & 0 & -0.508631 \\
%  0 & 0 & 0 & 0 & 0 & 0 & 1 & 0 \\
%  0 & 0.429338 & -0.504189 & 0 & -0.360084 & 0.288281 & 0 & 0.590505 \\
% \end{array}
% \right)
\,,
\end{split}
\end{equation}
and the superscript ``T'' denotes the transpose and the associated matrix is reported in Appendix~\ref{app:ccz_matrix}.\\
Figure~\ref{fig:CCZnew} and~\ref{fig:CCZnew_R} show such MZI networks embedded in $8\times 8$ Clements and Reck schemes, respectively, together with the MZIs settings.

Applying the linear transformation in Equation~\eqref{eq:recipeCCZ}, the state $\vert \Psi \rangle_{3,\rm in}$ evolves to $\vert \Psi \rangle_{3,\rm out}$, which reads modulo global phase as
\begin{equation}
    \begin{aligned}
        \vert \Psi \rangle_{3,\rm out}
        =&\, {\rm A}^{\rm CCZ}_{\rm succ} \left(   
        \alpha_{0} \alpha_{1} \alpha_{2} \,\hat{a}_{w_{0}^{0}}^\dagger \hat{a}_{w_{0}^{1}}^\dagger \hat{a}_{w_{0}^{2}}^\dagger + \alpha_{0}\alpha_{1} \beta_{2}\,\hat{a}_{w_{0}^{0}}^\dagger\hat{a}_{w_{0}^{1}}^\dagger \hat{a}_{w_{1}^{2}}^\dagger 
        + \alpha_{0} \beta_{1} \alpha_{2}\,\hat{a}_{w_{0}^{0}}^\dagger \hat{a}_{w_{1}^{1}}^\dagger \hat{a}_{w_{0}^{2}}^\dagger
        \right.\\
        & \left. \hspace{1cm}
        + \alpha_{0} \beta_{1} \beta_{2} \,\hat{a}_{w_{0}^{0}}^\dagger \hat{a}_{w_{1}^{1}}^\dagger \hat{a}_{w_{1}^{2}}^\dagger +
        \beta_{0} \alpha_{1} \alpha_{2}\,\hat{a}_{w_{1}^{0}}^\dagger \hat{a}_{w_{0}^{1}}^\dagger \hat{a}_{w_{0}^{2}}^\dagger + \beta_{0}\alpha_{1} \beta_{2}\,\hat{a}_{w_{1}^{0}}^\dagger\hat{a}_{w_{0}^{1}}^\dagger \hat{a}_{w_{1}^{2}}^\dagger
        \right.\\
        & \left. \hspace{1cm}
        + \beta_{0} \beta_{1} \alpha_{2}\,\hat{a}_{w_{1}^{0}}^\dagger \hat{a}_{w_{1}^{1}}^\dagger \hat{a}_{w_{0}^{2}}^\dagger
        - \beta_{0} \beta_{1} \beta_{2} \,\hat{a}_{w_{1}^{0}}^\dagger \hat{a}_{w_{1}^{1}}^\dagger \hat{a}_{w_{1}^{2}}^\dagger
        \right) \,\hat{a}_{w_{A}^{1}}^\dagger\, \hat{a}_{w_{A}^{2}}^\dagger \,\vert \Omega \rangle 
        +\ldots
    \end{aligned}
    \label{eq:ccz_hp}
\end{equation}
where ${\rm A}^{\rm CCZ}_{\rm succ} \approx 0.163231$ and the dots contain all the terms that do not preserve the initial qubit structure.

Finally, we can rewrite Equation~\eqref{eq:cz_hp} using the assignment with computational basis%, given Equation~\eqref{eq:basis}:
\begin{equation}
    \begin{aligned}
        \vert \Psi \rangle_{3,\rm out} \propto \,
        & \alpha_{0} \alpha_{1} \alpha_{2}\vert \mathsf{000} \rangle + \alpha_{0} \alpha_{1} \beta_{2} \vert \mathsf{001} \rangle + \alpha_{0} \beta_{1} \alpha_{2} \vert \mathsf{010} \rangle + \alpha_{0} \beta_{1} \beta_{2}\vert \mathsf{011} \rangle  \\
        & +\beta_{0} \alpha_{1} \alpha_{2} \vert \mathsf{100} \rangle + \beta_{0} \alpha_{1} \beta_{2} \vert \mathsf{101} \rangle + \beta_{0} \beta_{1} \alpha_{2} \vert \mathsf{110} \rangle - \beta_{0} \beta_{1} \beta_{2} \vert \mathsf{111} \rangle  \,,
    \end{aligned}
    \label{eq:ccz_corr}
\end{equation}
where we neglect the contributions that do not preserve the chosen qubit structure.

The success probability is 
\begin{equation}
    \mathbb{P}_{\rm succ}^{\rm CCZ} = \,\Big|{\rm A}^{\rm CCZ}_{\rm succ}\Big|^2 \approx 2.7 \% \,.
    \label{eq:ccz_succ}
\end{equation}

By taking $| q_0\rangle$ and $| q_1\rangle$  as the control qubits and $| q_2\rangle$ as the target qubit, CCNOT or Toffoli gate can be simply achieved by adding single-qubit gate as follows
\begin{equation}
    U_{8}^{\rm CCNOT} = 
     U_{78}(1/\sqrt{2}, 0) \cdot \bar{U}_8^{C/R}(\{ t_k^* , \phi_k^* \}_{k=1\ldots 10}) \cdot U_{78}(1/\sqrt{2}, 0) \,,
    \label{eq:ccnot_ps}
\end{equation}
and this gate can be easily embedded in both $8\times8$ schemes shown in Figure~\ref{fig:CCZnew} and ~\ref{fig:CCZnew_R} by changing the setting of two MZIs acting on modes $(7,8)$ before and after the central transformation.
It is possible to verify that the previous transformation produces the following output state
\begin{equation}
    \begin{aligned}
        \vert \Psi \rangle_{3,\rm out}
        =&\, {\rm A}^{\rm CCZ}_{\rm succ} \left(   
        \alpha_{0} \alpha_{1} \alpha_{2} \,\hat{a}_{w_{0}^{0}}^\dagger \hat{a}_{w_{0}^{1}}^\dagger \hat{a}_{w_{0}^{2}}^\dagger + \alpha_{0}\alpha_{1} \beta_{2}\,\hat{a}_{w_{0}^{0}}^\dagger\hat{a}_{w_{0}^{1}}^\dagger \hat{a}_{w_{1}^{2}}^\dagger 
        + \alpha_{0} \beta_{1} \alpha_{2}\,\hat{a}_{w_{0}^{0}}^\dagger \hat{a}_{w_{1}^{1}}^\dagger \hat{a}_{w_{0}^{2}}^\dagger
        \right.\\
        & \left. \hspace{1cm}
        + \alpha_{0} \beta_{1} \beta_{2} \,\hat{a}_{w_{0}^{0}}^\dagger \hat{a}_{w_{1}^{1}}^\dagger \hat{a}_{w_{1}^{2}}^\dagger +
        \beta_{0} \alpha_{1} \alpha_{2}\,\hat{a}_{w_{1}^{0}}^\dagger \hat{a}_{w_{0}^{1}}^\dagger \hat{a}_{w_{0}^{2}}^\dagger + \beta_{0}\alpha_{1} \beta_{2}\,\hat{a}_{w_{1}^{0}}^\dagger\hat{a}_{w_{0}^{1}}^\dagger \hat{a}_{w_{1}^{2}}^\dagger
        \right.\\
        & \left. \hspace{1cm}
        + \beta_{0} \beta_{1} \alpha_{2}\,\hat{a}_{w_{1}^{0}}^\dagger \hat{a}_{w_{1}^{1}}^\dagger \hat{a}_{w_{1}^{2}}^\dagger
        + \beta_{0} \beta_{1} \beta_{2} \,\hat{a}_{w_{1}^{0}}^\dagger \hat{a}_{w_{1}^{1}}^\dagger \hat{a}_{w_{0}^{2}}^\dagger
        \right) \,\hat{a}_{w_{A}^{1}}^\dagger\, \hat{a}_{w_{A}^{2}}^\dagger \,\vert \Omega \rangle 
        +\ldots
    \end{aligned}
    \label{eq:ccnot_hp}
\end{equation}
which is assigned to the correct output state
\begin{equation}
    \begin{aligned}
        \vert \Psi \rangle_{3,\rm out} \propto \,
        & \alpha_{0} \alpha_{1} \alpha_{2}\vert \mathsf{000} \rangle + \alpha_{0} \alpha_{1} \beta_{2} \vert \mathsf{001} \rangle + \alpha_{0} \beta_{1} \alpha_{2} \vert \mathsf{010} \rangle + \alpha_{0} \beta_{1} \beta_{2}\vert \mathsf{011} \rangle  \\
        & +\beta_{0} \alpha_{1} \alpha_{2} \vert \mathsf{100} \rangle + \beta_{0} \alpha_{1} \beta_{2} \vert \mathsf{101} \rangle + \beta_{0} \beta_{1} \alpha_{2} \vert \mathsf{111} \rangle + \beta_{0} \beta_{1} \beta_{2} \vert \mathsf{110} \rangle  \,,
    \end{aligned}
    \label{eq:ccnot_corr}
\end{equation}
when we consider only the output configurations preserving the qubit structure.

We conclude this section by pointing out that such post-selected CCZ gate cannot be cascaded on qubits' triplets that share at least two qubits. However, it is possible to cascade this operation on qubits' triplets that share only one qubit thanks to the presence of auxiliary photons. The details are reported in Section~\ref{app:ccz_matrix}.

\section{Conclusion}
\label{sec:conclu}

In this paper, we have shown how universal multiport interferometers can be used to implement universal two-qubit and three-qubit gates. The proposed approach can be also used for four-qubit gates following the same method used for the CZ and CCZ gates.
The qubit structure presented in Section~\ref{sec:qubitstruc} and shown in Figure~\ref{fig:new_scheme} offers a modification of standard dual-rail path encoding with improved performances. The proposed designs are feasible with respect to the current state of the art of single photon sources~\cite{ollivier2020reproducibility,suprano2023orbital} and integrated photonic architecture~\cite{bogaerts2018silicon,bogaerts2020programmable}. 
Even if the success probabilities are low, the presented gates can be used in shallow photonic hardware implementing variational quantum algorithms~\cite{Cerezo_2021}. These hybrid quantum-classical methods do not need error correction codes and many repetitions of entangling gates.

Comparing our results with those of~\cite{Kwon_24}, we obtain a higher success probability and the {\it truncation trick}, needed to cascade two CZ gates on two qubit pairs that share only one qubit, is avoided by adding one auxiliary photon in each auxiliary waveguide. Moreover, we have found designs for three-qubit gates without utilizing the composition of two-qubit gate designs.

\begin{table}[!h]
    \centering
    \begin{tabular}{c|c|c|c|c}
          % Proposed schemes 
          & Ancillary photons & Success probability & Feed-forward operation & Ref. \\ \hline
          \hline
          & two single photons & $1/16 \approx 6\%$ & no & \cite{knill_scheme_2001,Ralph_2001,okamoto2011realization}\\
          % & one Bell state  & $1/4 = 25\%$ & no & \cite{pittman2001probabilistic,pittman2002demonstration,gasparoni2004realization,zhao2005experimental}\\
          & no & $1/9 \approx 11.1\%$ & no & \cite{postsel_CZ,post_sel_2,obrien_demonstration_2003,langford2005demonstration,kiesel2005linear,okamoto2005demonstration,li2011reconfigurable,lee2022controlled}\\
          & two single photons & $1/8 = 12.5\%$ & no & \cite{bao2007optical,li2021heralded}\\
          & no & $1/8 = 12.5\%$ & no & \cite{liu2022universal} \\
          & a single photon & $1/4 = 25\%$ & yes & \cite{liu2023linear} \\
          & a single photon &  $\approx 15.2\%$ & no &  this work
    \end{tabular}
    \caption{Comparison of performances of optical CZ or CNOT gate.}
    \label{tab_comp_cz}
\end{table}

Table~\ref{tab_comp_cz} summarizes the comparison with other optical schemes for universal controlled gates, CZ and CNOT.
Table~\ref{tab_comp_ccz} presents the comparison with other optical universal controlled-controlled gates, CCZ and Toffoli.
Note that our gate designs for CZ and CCZ gates, have the second highest value of success probability and they are surpassed only by Ref.~\cite{liu2023linear}, where feed-forward operation is needed.
This operation is still demanding on an integrated platform. For example, if two gates with the feed-forward operation are executed, the hardware control requires the detection of the auxiliary photons from the first gate followed by a manipulation of the output photons fast enough to be performed before the action of the second gate. 

\begin{table}[!h]
    \centering
    \begin{tabular}{c|c|c|c|c}
          % Proposed schemes 
          & Ancillary photons & Success probability & Feed-forward operation & Ref. \\ \hline
          \hline
          & no & $1/133 \approx 0.8\%$ & no & \cite{fiuravsek2006linear}\\
          & no  & $1/72 = 1.4\%$ & no & \cite{ralph2007efficient,lanyon2009simplifying,li2022chip}\\
          % & two Bell states & $1/32 \approx 3.1\%$ & no & \cite{ralph2007efficient}\\
          & no & $1/64 = 1.6\%$ & no & \cite{liu2022universal} \\
          & no & $1/60 = 1.7\%$ & no & \cite{li2022quantum}\\
          & two single photons & $1/30 = 3.3\%$ & yes & \cite{liu2023linear} \\
          & two single photons &  $\approx 2.7\%$ & no &  this work
    \end{tabular}
    \caption{Comparison of performances of optical CCZ or Toffoli gate.}
    \label{tab_comp_ccz}
\end{table}

In these comparisons, we have not considered the CZ or CNOT gates that utilize one Bell state and have $1/4 = 25\%$ of success probability~\cite{pittman2001probabilistic,pittman2002demonstration,gasparoni2004realization,zhao2005experimental} and the CCZ or Toffoli gates based on two Bell states and reach $1/32 \approx 3.1\%$ of success probability~\cite{ralph2007efficient}. These optical gates reach the highest values of success probability, but they require initial auxiliary entangled states. Thus, with respect to the solutions reported in Table~\ref{tab_comp_cz} and Table~\ref{tab_comp_ccz}, these optical designs need more resources to execute the desired gates.

% Getting inspiration by observing that matrix permanents can be computed by Boson Sampling~\cite{boson_sampling} here we have applied Boson Sampling to post-selected multi-qubit gates.
Matrix permanent is the central concept of Boson Sampling~\cite{boson_sampling} and here we have used such non-trivial matrix property to build post-selected multi-qubit gates.
However, universal multiport interferometers have still richness and complexity to be investigated and we believe that the performances of post-selected multi-qubit gates can be improved by exploiting more involved structures.
For example, for the CZ gate we have simplified the problem by focusing on $3\times 3$ schemes embedded in $5\times 5$ schemes or for the CCZ gate we have looked for real solutions for $5\times 5$ schemes embedded in $8\times 8$ schemes.
As demonstrated in~\cite{bernstein1974must} and experimentally shown in~\cite{peruzzo2011multimode}, linear devices with more than three modes are described by an infinite number of distinct equivalence classes of transformations. This result suggests room for improvement.
Moreover, it is still unclear the upper bounds on the success probabilities for linear optical quantum computing gates.

Concerning scalability, our approach is characterized by some problems that must be solved or mitigated in order to reach fault-tolerant large-scale LOQC. 
Firstly, even in the ideal case, the success probability scales with multiples of about $15\%$ every time a post-selected CZ or CNOT gates and of about $2.7\%$ every time a post-selected CCZ or Toffoli gates are used.
Secondly, our scheme cannot implement QC circuits that involve the repetition of CZ and CNOT gates on the same qubits' pair and CCZ and Toffoli gates on triplets of qubits that share at least two qubits.
% This implies the impossibility of executing some quantum algorithms, like the Grover algorithm~\cite{samsonov2020modeling}.
From this perspective, the measurement-based approach~\cite{bartolucci_fusion-based_2023,briegel2001persistent,raussendorf2001one} through fusion gates has better performances on the integrated photonic platform. Therefore, the optimization of the success probability and the mitigation of the cascading issue are required for future developments of the gate-based approach in quantum photonics.
% The limitation in cascading multiple CZ on the same qubits' pair can be solved by modifying the high-level QC algorithms through additional ancillary qubits and gates in such a way as to avoid the cascading on the same pair of qubits. In this way, our scheme could be maintained for a generic quantum algorithm. Another non-conservative possibility consists of adopting another optical implementation of the CZ gate. For example, the KLM scheme or heralded CZ~\cite{knill_scheme_2001} allows the cascading in any configuration at the price of more resources and lower success probability. This choice would also imply the modification of the triplet structure in a new qubit structure.
% Finally, real integrated devices exhibit lower success probability and fidelity with respect to the ideal case: this holds not only for the probabilistic post-selected CZ gate, but also for deterministic gates like single-qubit and SWAP gates.
% Future investigations can also address the complete analysis of the performance and fidelity of the gates in the presence of real devices affected by non-idealities.'

Finally, the state-of-art of universal multiport interferometer experiments is the Gaussian Boson Sampling~\cite{hamilton2017gaussian,deng2023solving}. In Gaussian Boson Sampling continuous variable are used instead of discrete variable as in our approach. Moving from the discrete to continuous variables, for example by substituting single-photon states with single-mode squeezed vacuum, can provide another example of the use of post-selected photonic entangling gates. Indeed, the action of these can increase the non-triviality of the entangling resource processed in Gaussian Boson Sampling experiments.

\appendix
\begin{appendices}

\section{Integrated linear optical devices}
\label{app:preliminaries}

In this section, we give all the basic notions to describe linear optical devices and in particular we focus on the MZI action and its embedding in MZI networks.
The initial part follows the content of the appendix contained in \cite{Kwon_24}.

A device with $m$ inputs and $m$ outputs can be described by a $m\times m$ matrix $U$, 
\begin{equation}
    \begin{aligned}
    U =&   
    \begin{pmatrix}
        u_{11} & \ldots & u_{1m} \\
         \vdots &  \ddots &  \vdots \\
        u_{m1} & \ldots & u_{mm} \\
    \end{pmatrix} 
    \end{aligned}
    \label{unitarygen}
\end{equation}
where the components are complex number and $|\det(U)|=1$, if we have conservation of energy, or probability. 

Using the second quantization notation with creation and annihilation operators, we can define the input vector as $(a_1^\dagger,\ldots,a_m^\dagger)$ and the output vectors as $(b_1^\dagger,\ldots,b_m^\dagger)$, where we choose to count the modes from up position to down position.

The matrix $U$ establishes the relations between the inputs and outputs
\begin{align}
    \begin{pmatrix}
        b_1^\dagger \\
        \vdots \\
        b_m^\dagger
    \end{pmatrix}
    =
    \begin{pmatrix}
        u_{11} & \ldots & u_{1m} \\
         \vdots &  \ddots &  \vdots \\
        u_{m1} & \ldots & u_{mm} \\
    \end{pmatrix} 
    \begin{pmatrix}
        a_1^\dagger \\
        \vdots \\
        a_m^\dagger
    \end{pmatrix}
    = \begin{pmatrix}
        \sum_{k=1}^m u_{1k} \,a_k^\dagger \\
        \vdots \\
        \sum_{k=1}^m u_{mk} \,a_k^\dagger
    \end{pmatrix}
    \,.
\end{align}
Inverting the matrix $U$, we can invert the previous equation and use the relation from $a^\dagger$s to $b^\dagger$s to analyse the evolution of a given state:
\begin{equation}
    a^\dagger_k  \to \sum_{j=1}^m U^{-1}_{kj} a^\dagger_j \,,
    \label{eq:mxmgen}
\end{equation}
where $U^{-1}$ is the inverse of $U$.
% Let's make the example of $2\times2$ devices. In this case, we simply have 
% \begin{align}
%     \begin{pmatrix}
%         a_1^\dagger \\
%         a_2^\dagger
%     \end{pmatrix}
%     = \frac{1}{\det(U)}
%     \begin{pmatrix}
%         u_{22} & -u_{12} \\
%         -u_{21} & u_{11}
%     \end{pmatrix}
%     \begin{pmatrix}
%         b_1^\dagger \\
%         b_2^\dagger
%     \end{pmatrix}
%     = \frac{1}{\det(U)}
%     \begin{pmatrix}
%         u_{22} \,b_1^\dagger - u_{12}\,b_2^\dagger \\
%         -u_{21} \,b_1^\dagger + u_{11}\,b_2^\dagger
%     \end{pmatrix}\,.
%     \label{eq:2x2gen}
% \end{align}

Starting from the generic case, it is possible to construct any unitary matrix $m\times m$ from $2\times2$ sub-matrices. Concretely, this is equivalent to creating a generic linear $m\times m$ device assembling linear $2\times2$ devices, like MZI.
Indeed, the Reck and Clements~\cite{reck_experimental_1994,clements_optimal_2016} schemes give exactly two prescriptions to obtain a generic unitary operation, described in the transformation~\eqref{unitarygen}, by assembling MZIs.

In order to describe the assembling, we define a generic embedded $2\times2$ matrix $U$ in $m\times m$ system. The transformation of the $k$-th and $(k+1)$-th inputs reads 
\begin{equation}
U^{(k,k+1)} \equiv
    \begin{pmatrix}
    1 & 0 &  \ldots & \ldots & \ldots & 0 \\
    0 & \ddots & & & &\vdots \\
    \vdots & & u_{11} & u_{12} & & \vdots \\
    \vdots & & u_{21} & u_{22} & & \vdots \\
    \vdots & & & & \ddots & 0 \\
    0 & \ldots & \ldots & \ldots & 0 & 1 \\
    \end{pmatrix} \,.
    \label{Ukemb}
\end{equation}
This transformation is leaving unaffected all inputs different from $k$ and $(k+1)$ ones, which are evolving accordingly to Equation~\eqref{eq:mxmgen}.

\begin{figure}
    \centering
    \includegraphics[width=\linewidth]{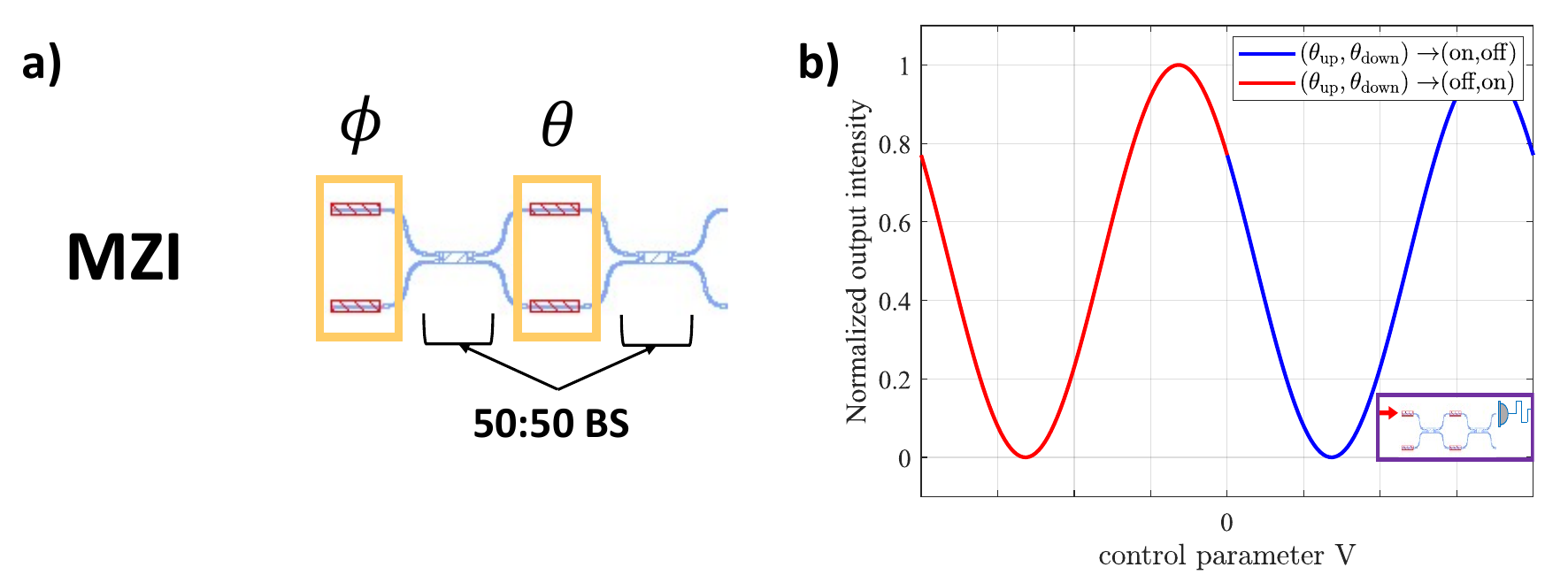}
    \caption{ (a) Representation of the MZI unit. Light blue parts are waveguides and integrated beam splitters and red parts are phase shifters.
    This device has two inputs and two outputs and is composed of 4 phase shifters and two balanced integrated beam splitters. 
    This configuration with pairs of phase shifters is called "push\&pull", but the MZI unit can be built also with half of the phase shifters.
   (b) Example of output optical intensity at one of the two outputs of the MZI when light is injected in only one input. The blue/red line is obtained by using the upper/lower phase shifter, denoted by $\theta$. Note that we insert a spurious phase in the internal arms, since with the control parameter $V$ equal to zero the output normalized intensity is not either one or zero.
    }
    \label{fig:MZI}
\end{figure}

At this point, we show how to construct the MZI: the ingredients are two balanced beam splitters (BSs) and two pairs of phase shifters (PSs).
Balanced ideal BS can be represented with the following matrix
\begin{equation}
    {\rm BS} = \frac{1}{\sqrt{2}}
    \begin{pmatrix}
        1 & \pm{\rm i} \\
        \pm{\rm i} & 1
    \end{pmatrix} \,,
    \label{eq:beam_ideal}
\end{equation}
modulo a global phase. The plus sign corresponds to the MMIs case, while the minus sign to the directional coupler case. Along the manuscript for simplicity, we decide to use the plus sign version without loss of generality. The previous matrix can be compared to the case of BS with general transmittance
\begin{equation}
    {\rm BS}_{\rm gen}(t,r) =
    \begin{pmatrix}
        t & {\rm i}\,r \\
        {\rm i}\,r & t
    \end{pmatrix} \,,
    \label{eq:beam_general}
 \end{equation}
where the label ``gen'' stands for generic and $t$ and $r$ are the amplitudes of the two outputs and they are denoted as transmittance and reflectance. Unbalance means that $|t| \ne |r|$ and insertion loss implies that  $|t|^2 + |r|^2 < 1$.
For a pair of waveguides, PSs on each path introduce two different phases. This configuration involves two inputs and two outputs and it can be described by the following matrix
\begin{equation}
    {\rm PS}(\boldsymbol{\theta}) = 
    \begin{pmatrix}
        {\rm e}^{{\rm i}\theta_1} & 0 \\
        0 & {\rm e}^{{\rm i}\theta_2}
    \end{pmatrix} \,,
    \label{eq:phase_shift}
\end{equation}
where $\boldsymbol{\theta} = (\theta_1,\theta_2)$.
Note that after this operation, we can measure only the relative phase $\theta_1-\theta_2$, if we look just at one pair of waveguides. 

Consequently, the unitary matrix for a MZI is 
\begin{equation}
    \begin{split}
    U_{\rm MZI}( \boldsymbol{\theta},\boldsymbol{\phi} )  &\equiv {\rm BS} \cdot {\rm PS}(\boldsymbol{\theta}) \cdot {\rm BS} \cdot {\rm PS}(\boldsymbol{\phi}) \\
    &= {\rm i}\, {\rm e}^{{\rm i}(\theta_1+\theta_2+2\phi_2)/2} 
    ~~~~
    \begin{pmatrix}
        {\rm e}^{{\rm i}(\phi_1-\phi_2)}\sin((\theta_1-\theta_2)/2) & \cos((\theta_1-\theta_2)/2) \\
        {\rm e}^{{\rm i}(\phi_1-\phi_2)}\cos((\theta_1-\theta_2)/2) & -\sin((\theta_1-\theta_2)/2)
    \end{pmatrix} \,,
    \end{split}
\label{eq:MZI_matrixapp}
\end{equation}
where $\boldsymbol{\theta} = (\theta_1,\theta_2)$ and $\boldsymbol{\phi} = (\phi_1,\phi_2)$.
Note that if we define the transmittance $t \equiv \sin((\theta_1-\theta_2)/2)$, Equation~\eqref{eq:MZI_matrix} has the same representation as Equation~\eqref{eq:beam_general}, which describes a generic beam splitter. This means that the MZI is equivalent to a beam splitter with tunable transmittance.
Negative transmittance values mean that light is transmitted with a ratio given by the squared value of the transmittance and the reflected and transmitted portions of light have an additional $\pi$ shift difference.
As we said in the case of PS, we can measure only the relative phase $\theta_1-\theta_2$ and $\phi_1-\phi_2$, if we look just at the specific pair of waveguides connected to the MZI.

Figure~\ref{fig:MZI}(a) shows an integrated MZI in the configuration with two phase shifters to tune $\boldsymbol{\theta}$ and two for $\boldsymbol{\phi}$, called "push\&pull". It is also possible to work with only one phase shifter for $\boldsymbol{\theta}$ and one for $\boldsymbol{\phi}$, but the "push\&pull" configuration typically has advantages regarding power consumption. In both cases, $\boldsymbol{\theta}$ and $\boldsymbol{\phi}$ are functions of an external parameter $V$ that we can vary to reconfigure the MZI to the desired configuration. The action of $V$ produces an additional phase with respect to the intrinsic phase given by the propagation.
Figure~\ref{fig:MZI}(b) shows an example of the output normalized power for an integrated "push\&pull" MZI when light is injected in only one input and we vary either $\theta_1$ or $\theta_2$. The plot shows a value of output power for zero control parameter $V$ that is not either one or zero: this is due to the presence of a spurious phase given by the finite resolution of the fabrication process.
Note that we can actively decide to explore more easily the different output configurations through the tuning of either $\theta_1$ or $\theta_2$. 
In general, in order to obtain $t=\sin \Theta/2$ with $\Theta \in (-2\pi,2\pi)$, one can tune 
\begin{equation}
    \theta_1(V) \to 
    \begin{cases}
        \theta_2(0)+\Theta & \mbox{if}\quad \Theta>0\,, \\
        \theta_2(0)+2\pi+\Theta & \mbox{if}\quad \Theta<0\,,
    \end{cases}
\end{equation} or 
\begin{equation}
    \theta_2(V) \to 
    \begin{cases}
        \theta_1(0)+2\pi-\Theta & \mbox{if}\quad \Theta>0\,, \\
        \theta_1(0)-\Theta & \mbox{if}\quad \Theta<0\,.
    \end{cases}
\end{equation}
For example, if $t=\pm 1/\sqrt{2}$ or $\Theta=\pm \pi/2$, $\theta_1(V)$ is set to be $\theta_2(0)+\pi/2$ and $\theta_2(0)+3\pi/2$, respectively, or $\theta_2(V)$ to $\theta_1(0)+3\pi/2$ and $\theta_1(0)+\pi/2$, respectively.
The same logic holds for phase shifters associated with $\phi_1$ and $\phi_2$.

\section{Controlled gates}
\label{app:controlledgates}
In this section, we report logic tables for Z, X, Controlled-Z, Controlled-NOT, Controlled-Controlled-Z and Toffoli gates, Tables~\ref{Ztable}-\ref{CZtable}-\ref{CCZtable}.

Z and X gates are single-qubit transformations: the first one imposes $\pi$ phase shift only if the qubit is in state $|1\rangle$, while the second one acts as NOT gate on the basis states.
\begin{table}[!h]
    \centering
    \begin{tabular}{|c||c||c|}
         \hline
          input state & Z output state & X output state\\ 
          \hline
          \hline
          $|0\rangle$ & $|0\rangle$ & $|1\rangle$ \\
          $|1\rangle$ & $-|1\rangle$ & $|0\rangle$ \\
          \hline
    \end{tabular}
    \caption{Logic tables for the Z and X gates.}
    \label{Ztable}
\end{table}

In the case of CZ and CNOT gates, the first qubit is the controlled qubit and the second one is the target qubit. These two-qubit gates execute Z gate or X gate respectively on the target qubit only when the control qubit is in state $|1\rangle$ and no action otherwise.

\begin{table}[!h]
    \centering
    \begin{tabular}{|c||c||c|}
         \hline
          input state & CZ output state & CNOT output state\\ 
          \hline
          \hline
          $|0\rangle \otimes |0\rangle$ & $|0\rangle \otimes |0\rangle$ & $|0\rangle \otimes |0\rangle$ \\
          $|0\rangle \otimes |1\rangle$ & $|0\rangle \otimes |1\rangle$ & $|0\rangle \otimes |1\rangle$ \\
          $|1\rangle \otimes |0\rangle$ & $|1\rangle \otimes |0\rangle$ & $|1\rangle \otimes |1\rangle$ \\
          $|1\rangle \otimes |1\rangle$ & $-|1\rangle \otimes |1\rangle$ & $|1\rangle \otimes |0\rangle$ \\
          \hline
    \end{tabular}
    \caption{Logic tables for the CZ and CNOT gates.}
    \label{CZtable}
\end{table}

In the case of CCZ and Toffoli gates, the first and second qubits are the controlled qubits and the third one is the target qubit. These three-qubit gates execute Z gate or X gate respectively on the target qubit only when both control qubits are in state $|1\rangle$ and no action otherwise.
\begin{table}[!h]
    \centering
    \begin{tabular}{|c||c||c|}
         \hline
          input state & CCZ output state & Toffoli output state\\ 
          \hline
          \hline
          $|0\rangle \otimes |0\rangle \otimes |0\rangle$  & $|0\rangle \otimes |0\rangle \otimes |0\rangle$  & $|0\rangle \otimes |0\rangle \otimes |0\rangle$ \\
          $|0\rangle \otimes |0\rangle \otimes |1\rangle$  & $|0\rangle \otimes |0\rangle \otimes |1\rangle$  & $|0\rangle \otimes |0\rangle \otimes |1\rangle$ \\
          $|0\rangle \otimes |1\rangle \otimes |0\rangle$  & $|0\rangle \otimes |1\rangle \otimes |0\rangle$  & $|0\rangle \otimes |1\rangle \otimes |0\rangle$ \\
          $|0\rangle \otimes |1\rangle \otimes |1\rangle$  & $|0\rangle \otimes |1\rangle \otimes |1\rangle$  & $|0\rangle \otimes |1\rangle \otimes |1\rangle$ \\
          $|1\rangle \otimes |0\rangle \otimes |0\rangle$  & $|1\rangle \otimes |0\rangle \otimes |0\rangle$  & $|1\rangle \otimes |0\rangle \otimes |0\rangle$ \\
          $|1\rangle \otimes |0\rangle \otimes |1\rangle$  & $|1\rangle \otimes |0\rangle \otimes |1\rangle$  & $|1\rangle \otimes |0\rangle \otimes |1\rangle$ \\
          $|1\rangle \otimes |1\rangle \otimes |0\rangle$  & $|1\rangle \otimes |1\rangle \otimes |0\rangle$  & $|1\rangle \otimes |1\rangle \otimes |1\rangle$ \\
          $|1\rangle \otimes |1\rangle \otimes |1\rangle$  & $-|1\rangle \otimes |1\rangle \otimes |1\rangle$  & $|1\rangle \otimes |1\rangle \otimes |0\rangle$ \\
          \hline
    \end{tabular}
    \caption{Logic tables for the CCZ and Toffoli gates.}
    \label{CCZtable}
\end{table}

\section{How to cascade post-selected CZ }
\label{app:cascade}

In this section, we show that the post-selected CZ presented in Sec.~\ref{sec:cz} can be cascaded to two pairs of qubits that share only one qubit.
Thus, let's consider three qubits as in Eq.~\eqref{eq:generic_3qstate}.
As in the main text, we arrange the creation operators in a vector as follows
\begin{equation}
    {\rm V}_3^{\rm T} \equiv 
    \left(
        \hat{a}_{w_{0}^{0}}^\dagger,
        \hat{a}_{w_{1}^{0}}^\dagger,
        \hat{a}_{w_{A}^{1}}^\dagger,
        \hat{a}_{w_{0}^{1}}^\dagger,
        \hat{a}_{w_{1}^{1}}^\dagger,
        \hat{a}_{w_{A}^{2}}^\dagger,
        \hat{a}_{w_{0}^{2}}^\dagger,
        \hat{a}_{w_{1}^{2}}^\dagger
    \right) \,.
\end{equation}
If we apply the post-selected CZ firstly to the first and second qubits and then to the second and third qubits as 
\begin{equation}
    {\rm V}_3 \to \left( U_{5}^{\rm CZ}(2,3) \right)^{-1} \cdot \left( U_{5}^{\rm CZ}(1,2) \right)^{-1} \cdot {\rm V}_3 \,,
\end{equation}
where
\begin{equation}
\begin{split}
    U_{5}^{\rm CZ}(1,2) &\equiv 
    \footnotesize
    \left(
    \begin{array}{cccccccc}
     1 & 0 & 0 & 0 & 0 & 0 & 0 & 0 \\
     0 & 0.2192 & 0.8475 & 0 & 0.4834 & 0 & 0 & 0 \\
     0 & 0.3597 & 0.3904 & 0 & -0.8475 & 0 & 0 & 0 \\
     0 & 0 & 0 & 1 & 0 & 0 & 0 & 0 \\
     0 & 0.9070 & -0.3597 & 0 & 0.2192 & 0 & 0 & 0 \\
     0 & 0 & 0 & 0 & 0 & 1 & 0 & 0 \\
     0 & 0 & 0 & 0 & 0 & 0 & 1 & 0 \\
     0 & 0 & 0 & 0 & 0 & 0 & 0 & 1 \\
    \end{array}
    \right)\,,\\
    U_{5}^{\rm CZ}(2,3) &\equiv
    \footnotesize
    \left(
    \begin{array}{cccccccc}
     1 & 0 & 0 & 0 & 0 & 0 & 0 & 0 \\
     0 & 1 & 0 & 0 & 0 & 0 & 0 & 0 \\
     0 & 0 & 1 & 0 & 0 & 0 & 0 & 0 \\
     0 & 0 & 0 & 1 & 0 & 0 & 0 & 0 \\
     0 & 0 & 0 & 0 & 0.2192 & 0.8475 & 0 & 0.4834 \\
     0 & 0 & 0 & 0 & 0.3597 & 0.3904 & 0 & -0.8475 \\
     0 & 0 & 0 & 0 & 0 & 0 & 1 & 0 \\
     0 & 0 & 0 & 0 & 0.9070 & -0.3597 & 0 & 0.2192 \\
    \end{array}
    \right)\,.
\end{split}
\end{equation}

It is possible to verify that given the generic initial state,
\begin{equation}
    \vert \Psi_3 \rangle_{\rm in} = \left( A_{0} \,\hat{a}_{w_{0}^{0}}^\dagger + A_{1} \,\hat{a}_{w_{1}^{0}}^\dagger \right)
    \,\hat{a}_{w_{A}^1}^\dagger \,
    \left( B_{0} \,\hat{a}_{w_{0}^{1}}^\dagger + B_{1} \,\hat{a}_{w_{1}^{1}}^\dagger \right) 
    \,\hat{a}_{w_{A}^2}^\dagger \,
    \left( C_{0} \,\hat{a}_{w_{0}^{2}}^\dagger + C_{1} \,\hat{a}_{w_{1}^{2}}^\dagger \right)
    \vert \Omega \rangle\,
\end{equation}
the correct output state is obtained,
\begin{equation}
    \vert \Psi_3 \rangle_{\rm out} =
    % \, {\rm CZ}_{\rm ps}^{(2,3)} \cdot \hat{\rm P}_{\rm aux} \cdot {\rm CZ}_{\rm ps}^{(1,2)} \, \vert \Psi_3 \rangle_{\rm ini} \\
    % & = \frac{1}{9}\left(  
    % A_0 B_0 C_0 \,\hat{a}_{0}^\dagger \hat{b}_{0}^\dagger \hat{c}_{0}^\dagger 
    % + A_0 B_0 C_1\,\hat{a}_{0}^\dagger\hat{b}_{0}^\dagger \hat{c}_{1}^\dagger
    % + A_0 B_1 C_0\,\hat{a}_{0}^\dagger\hat{b}_{1}^\dagger \hat{c}_{0}^\dagger 
    % - A_0 B_1 C_1\,\hat{a}_{0}^\dagger\hat{b}_{1}^\dagger \hat{c}_{1}^\dagger 
    % \right.\\
    % &\hspace{1cm}\left.+ A_1 B_0 C_0\,\hat{a}_{1}^\dagger\hat{b}_{0}^\dagger \hat{c}_{0}^\dagger 
    % + A_1 B_0 C_1\,\hat{a}_{1}^\dagger\hat{b}_{0}^\dagger \hat{c}_{1}^\dagger 
    % - A_1 B_1 C_0\,\hat{a}_{1}^\dagger\hat{b}_{1}^\dagger \hat{c}_{0}^\dagger 
    % + A_1 B_1 C_1\,\hat{a}_{1}^\dagger\hat{b}_{1}^\dagger \hat{c}_{1}^\dagger
    % \right) \vert \Omega \rangle +\ldots \\
    % &= 
    \left({\rm A}^{\rm CZ}_{\rm succ}\right)^2 \,\hat{a}_{w_{A}^1}^\dagger \,\hat{a}_{w_{A}^2}^\dagger \,\sum_{j_0,j_1,j_2=\{0,1\}} (-)^{j_0\,j_1+j_1\,j_2}\,A_{j_0} B_{j_1} C_{j_2} \,\hat{a}_{w^0_{j_0}}^\dagger \hat{a}_{w^1_{j_1}}^\dagger\hat{a}_{w^2_{j_2}}^\dagger \,\vert \Omega \rangle
        +\ldots\,,
\end{equation}
where the dots contain all the terms that do not preserve the qubit structure. Indeed, we achieve the following desired evolution
\begin{equation}
    \vert q_0 \rangle_{\rm in} \otimes \vert q_1 \rangle_{\rm in} \otimes \vert q_2 \rangle_{\rm in} 
    \to \sum_{j_0,j_1,j_2=\{0,1\}} (-)^{j_0\,j_1+j_1\,j_2}\,A_{j_0} B_{j_1} C_{j_2} \,\vert j_0 \rangle \otimes \vert j_1 \rangle \otimes \vert j_2 \rangle \,,
\end{equation}
where $\vert q_0 \rangle_{\rm in} = \left( A_{0} \,\vert \mathsf{0} \rangle + A_{1} \,\vert \mathsf{1} \rangle \right)$, $\vert q_1 \rangle_{\rm in} = \left( B_{0} \,\vert \mathsf{0} \rangle + B_{1} \,\vert \mathsf{1} \rangle \right)$ and $\vert q_2 \rangle_{\rm in} = \left( C_{0} \,\vert \mathsf{0} \rangle + C_{1} \,\vert \mathsf{1} \rangle \right)$.

\section{Post-selected CZ with more auxiliary single photons}
\label{app:auxtower}

In this section, we summarized the cases where we modify the qubit structure presented in Fig.~\ref{fig:new_scheme} adding more auxiliary single photons in each auxiliary path.
It is possible to find a tower of solutions as the number of auxiliary photons increases. These solutions are listed in Table~\ref{more_aux_pho}.
% \begin{equation}
% \begin{aligned}
%     & \mbox{2 auxiliary photons} 
%     \begin{cases}
%         & {\rm A}^{\rm CZ}_{\rm succ} \approx 0.310137 \,,\\
%         & (t^*_1,t^*_2,t^*_3) = (0.309466, -0.151715, 0.781247) \,, 
%     \end{cases}\\
%     & \mbox{3 auxiliary photons} 
%     \begin{cases}
%         & {\rm A}^{\rm CZ}_{\rm succ} \approx  0.281103\,,\\
%         & (t^*_1,t^*_2,t^*_3) = (0.275836, -0.116555, 0.708984) \,, 
%     \end{cases}\\
%     & \mbox{4 auxiliary photons} 
%     \begin{cases}
%         & {\rm A}^{\rm CZ}_{\rm succ} \approx  0.266422\,,\\
%         & (t^*_1,t^*_2,t^*_3) = (0.251741, -0.0946289, 0.651796) \,, 
%     \end{cases}\\
%     & \mbox{5 auxiliary photons} 
%     \begin{cases}
%         & {\rm A}^{\rm CZ}_{\rm succ} \approx  0.25762\,,\\
%         & (t^*_1,t^*_2,t^*_3) = (0.23313, -0.0796303, 0.605833) \,, 
%     \end{cases}\\
%     & \mbox{6 auxiliary photons} 
%     \begin{cases}
%         & {\rm A}^{\rm CZ}_{\rm succ} \approx  0.251772\,,\\
%         & (t^*_1,t^*_2,t^*_3) = (0.218142, -0.0687255, 0.568082) \,, 
%     \end{cases}\\
%     & \mbox{7 auxiliary photons} 
%     \begin{cases}
%         & {\rm A}^{\rm CZ}_{\rm succ} \approx 0.247611\,,\\
%         & (t^*_1,t^*_2,t^*_3) = (0.205723, -0.0604419, 0.536449) \,, 
%     \end{cases}\\
%     \ldots
% \end{aligned}
% \end{equation}

\begin{table}[!h]
    \centering
    \begin{tabular}{c|c|c|c}
          % Proposed schemes 
          & Ancillary photons & t-parameters: $( t^*_1,t^*_2,t^*_3)$ & ${\rm A}^{\rm CZ}_{\rm succ}$ \\ \hline
          \hline
          & one & $( 0.3686, -0.2192, 0.8686 ) $ & $0.3904$ \\
          & two  & $(0.3095, -0.1517, 0.7812)$ & $0.3101$ \\
          & three & $(0.2758, -0.1166, 0.7090)$ & $0.2811$ \\
          & four & $(0.2517, -0.09463, 0.6518)$ & $0.2664$ \\
          & five & $(0.2331, -0.07963, 0.6058)$ & $0.2576$ \\
          & six & $(0.2181, -0.06873, 0.5681)$ & $0.2518$ \\
          & seven &  $(0.2057, -0.06044, 0.5364)$ & $0.2476$ \\
          & $\ldots$ &  $\ldots$ & $\ldots$
    \end{tabular}
    \caption{Tower of solutions as we increase the number of auxiliary single photons in the dual-rail path encoding with auxiliary paths, presented in Fig.~\ref{fig:new_scheme}.}
    \label{more_aux_pho}
\end{table}

A decreasing pattern appears for the success probability as the number of auxiliary photons is increased: thus, there is no reason to add them to the qubit structure shown in Fig.~\ref{fig:new_scheme}.

\section{Details about MZI network for post-selected CCZ gate}
\label{app:ccz_matrix}

In this section, we present the matrix associated with both $8\times 8$ Clements and Reck schemes shown in Fig.~\ref{fig:CCZnew} and Fig~\ref{fig:CCZnew_R}.

The matrix is the same but decomposed in two different ways presented in Eq.~\eqref{eq:clem8} and Eq.~\eqref{eq:reck8}, and it reads
\begin{equation}
    U_{8}^{\rm CCZ} \equiv 
\left(
\begin{array}{cccccccc}
 1. & 0. & 0. & 0. & 0. & 0. & 0. & 0. \\
 0. & -0.2531 & -0.1447 & 0. & -0.2891 & -0.8597 & 0. & 0.3039 \\
 0. & -0.0950 & -0.3670 & 0. & 0.8626 & -0.0859 & 0. & 0.3236 \\
 0. & 0. & 0. & 1. & 0. & 0. & 0. & 0. \\
 0. & -0.4875 & -0.7133 & 0. & -0.1747 & 0.1660 & 0. & -0.4421 \\
 0. & 0.7106 & -0.2853 & 0. & 0.1100 & -0.3780 & 0. & -0.5086 \\
 0. & 0. & 0. & 0. & 0. & 0. & 1. & 0. \\
 0. & 0.4293 & -0.5042 & 0. & -0.3601 & 0.2883 & 0. & 0.5905 \\
\end{array}
\right) \,.
\end{equation}

We conclude this section by showing that the post-selected CCZ presented in Sec.~\ref{sec:ccz} can be cascaded to two triplets of qubits that share only one qubit.
Thus, let's consider five qubits.
As in the main text, we arrange the creation operators in a vector as follows
\begin{equation}
    {\rm V}_5^{\rm T} \equiv 
    \left(
        \hat{a}_{w_{0}^{0}}^\dagger,
        \hat{a}_{w_{1}^{0}}^\dagger,
        \hat{a}_{w_{A}^{1}}^\dagger,
        \hat{a}_{w_{0}^{1}}^\dagger,
        \hat{a}_{w_{1}^{1}}^\dagger,
        \hat{a}_{w_{A}^{2}}^\dagger,
        \hat{a}_{w_{0}^{2}}^\dagger,
        \hat{a}_{w_{1}^{2}}^\dagger,
        \hat{a}_{w_{A}^{3}}^\dagger,
        \hat{a}_{w_{0}^{3}}^\dagger,
        \hat{a}_{w_{1}^{3}}^\dagger,
        \hat{a}_{w_{A}^{4}}^\dagger,
        \hat{a}_{w_{0}^{4}}^\dagger,
        \hat{a}_{w_{1}^{4}}^\dagger
    \right) \,.
\end{equation}
If we apply the post-selected CCZ firstly to the first, second and third qubits and then to the third, fourth and fifth qubits as 
\begin{equation}
    {\rm V}_5 \to \left( U_{8}^{\rm CCZ}(3,4,5) \right)^{-1} \cdot \left( U_{8}^{\rm CCZ}(1,2,3) \right)^{-1} \cdot {\rm V}_5 \,,
\end{equation}
where
\begin{equation}
\begin{split}
    U_{8}^{\rm CCZ}(1,2,3) &\equiv 
    \footnotesize
    \left(
    \begin{array}{cccccccccccccc}
     1. & 0. & 0. & 0. & 0. & 0. & 0. & 0. & 0. & 0. & 0. & 0. & 0. & 0. \\
     0. & -0.253097 & -0.144686 & 0. & -0.289053 & -0.859702 & 0. & 0.303922 & 0. & 0. & 0. & 0. & 0. & 0. \\
     0. & -0.0949594 & -0.367007 & 0. & 0.862646 & -0.0859038 & 0. & 0.323652 & 0. & 0. & 0. & 0. & 0. & 0. \\
     0. & 0. & 0. & 1. & 0. & 0. & 0. & 0. & 0. & 0. & 0. & 0. & 0. & 0. \\
     0. & -0.487524 & -0.713273 & 0. & -0.174711 & 0.16601 & 0. & -0.44213 & 0. & 0. & 0. & 0. & 0. & 0. \\
     0. & 0.710573 & -0.285319 & 0. & 0.110023 & -0.377978 & 0. & -0.508631 & 0. & 0. & 0. & 0. & 0. & 0. \\
     0. & 0. & 0. & 0. & 0. & 0. & 1. & 0. & 0. & 0. & 0. & 0. & 0. & 0. \\
     0. & 0.429338 & -0.504189 & 0. & -0.360084 & 0.288281 & 0. & 0.590505 & 0. & 0. & 0. & 0. & 0. & 0. \\
     0. & 0. & 0. & 0. & 0. & 0. & 0. & 0. & 1. & 0. & 0. & 0. & 0. & 0. \\
     0. & 0. & 0. & 0. & 0. & 0. & 0. & 0. & 0. & 1. & 0. & 0. & 0. & 0. \\
     0. & 0. & 0. & 0. & 0. & 0. & 0. & 0. & 0. & 0. & 1. & 0. & 0. & 0. \\
     0. & 0. & 0. & 0. & 0. & 0. & 0. & 0. & 0. & 0. & 0. & 1. & 0. & 0. \\
     0. & 0. & 0. & 0. & 0. & 0. & 0. & 0. & 0. & 0. & 0. & 0. & 1. & 0. \\
     0. & 0. & 0. & 0. & 0. & 0. & 0. & 0. & 0. & 0. & 0. & 0. & 0. & 1. \\
    \end{array}
    \right)\,,\\
    U_{8}^{\rm CCZ}(3,4,5) &\equiv
    \footnotesize
    \left(
    \begin{array}{cccccccccccccc}
     1. & 0. & 0. & 0. & 0. & 0. & 0. & 0. & 0. & 0. & 0. & 0. & 0. & 0. \\
     0. & 1. & 0. & 0. & 0. & 0. & 0. & 0. & 0. & 0. & 0. & 0. & 0. & 0. \\
     0. & 0. & 1. & 0. & 0. & 0. & 0. & 0. & 0. & 0. & 0. & 0. & 0. & 0. \\
     0. & 0. & 0. & 1. & 0. & 0. & 0. & 0. & 0. & 0. & 0. & 0. & 0. & 0. \\
     0. & 0. & 0. & 0. & 1. & 0. & 0. & 0. & 0. & 0. & 0. & 0. & 0. & 0. \\
     0. & 0. & 0. & 0. & 0. & 1. & 0. & 0. & 0. & 0. & 0. & 0. & 0. & 0. \\
     0. & 0. & 0. & 0. & 0. & 0. & 1. & 0. & 0. & 0. & 0. & 0. & 0. & 0. \\
     0. & 0. & 0. & 0. & 0. & 0. & 0. & -0.253097 & -0.144686 & 0. & -0.289053 & -0.859702 & 0. & 0.303922 \\
     0. & 0. & 0. & 0. & 0. & 0. & 0. & -0.0949594 & -0.367007 & 0. & 0.862646 & -0.0859038 & 0. & 0.323652 \\
     0. & 0. & 0. & 0. & 0. & 0. & 0. & 0. & 0. & 1. & 0. & 0. & 0. & 0. \\
     0. & 0. & 0. & 0. & 0. & 0. & 0. & -0.487524 & -0.713273 & 0. & -0.174711 & 0.16601 & 0. & -0.44213 \\
     0. & 0. & 0. & 0. & 0. & 0. & 0. & 0.710573 & -0.285319 & 0. & 0.110023 & -0.377978 & 0. & -0.508631 \\
     0. & 0. & 0. & 0. & 0. & 0. & 0. & 0. & 0. & 0. & 0. & 0. & 1. & 0. \\
     0. & 0. & 0. & 0. & 0. & 0. & 0. & 0.429338 & -0.504189 & 0. & -0.360084 & 0.288281 & 0. & 0.590505 \\
    \end{array}
    \right)\,.
\end{split}
\end{equation}

It is possible to verify that given the generic initial state,
\begin{equation}
\footnotesize
    \vert \Psi_5 \rangle_{\rm in} = \left( A_{0} \,\hat{a}_{w_{0}^{0}}^\dagger + A_{1} \,\hat{a}_{w_{1}^{0}}^\dagger \right)
    \,\hat{a}_{w_{A}^1}^\dagger \,
    \left( B_{0} \,\hat{a}_{w_{0}^{1}}^\dagger + B_{1} \,\hat{a}_{w_{1}^{1}}^\dagger \right) 
    \,\hat{a}_{w_{A}^2}^\dagger \,
    \left( C_{0} \,\hat{a}_{w_{0}^{2}}^\dagger + C_{1} \,\hat{a}_{w_{1}^{2}}^\dagger \right)
    \,\hat{a}_{w_{A}^3}^\dagger \,
    \left( D_{0} \,\hat{a}_{w_{0}^{3}}^\dagger + D_{1} \,\hat{a}_{w_{1}^{3}}^\dagger \right)
    \,\hat{a}_{w_{A}^4}^\dagger \,
    \left( E_{0} \,\hat{a}_{w_{0}^{4}}^\dagger + E_{1} \,\hat{a}_{w_{1}^{4}}^\dagger \right)
    \vert \Omega \rangle\,
\end{equation}
the correct output state is obtained,
\begin{equation}
    \vert \Psi_5 \rangle_{\rm out} =
    % \, {\rm CZ}_{\rm ps}^{(2,3)} \cdot \hat{\rm P}_{\rm aux} \cdot {\rm CZ}_{\rm ps}^{(1,2)} \, \vert \Psi_3 \rangle_{\rm ini} \\
    % & = \frac{1}{9}\left(  
    % A_0 B_0 C_0 \,\hat{a}_{0}^\dagger \hat{b}_{0}^\dagger \hat{c}_{0}^\dagger 
    % + A_0 B_0 C_1\,\hat{a}_{0}^\dagger\hat{b}_{0}^\dagger \hat{c}_{1}^\dagger
    % + A_0 B_1 C_0\,\hat{a}_{0}^\dagger\hat{b}_{1}^\dagger \hat{c}_{0}^\dagger 
    % - A_0 B_1 C_1\,\hat{a}_{0}^\dagger\hat{b}_{1}^\dagger \hat{c}_{1}^\dagger 
    % \right.\\
    % &\hspace{1cm}\left.+ A_1 B_0 C_0\,\hat{a}_{1}^\dagger\hat{b}_{0}^\dagger \hat{c}_{0}^\dagger 
    % + A_1 B_0 C_1\,\hat{a}_{1}^\dagger\hat{b}_{0}^\dagger \hat{c}_{1}^\dagger 
    % - A_1 B_1 C_0\,\hat{a}_{1}^\dagger\hat{b}_{1}^\dagger \hat{c}_{0}^\dagger 
    % + A_1 B_1 C_1\,\hat{a}_{1}^\dagger\hat{b}_{1}^\dagger \hat{c}_{1}^\dagger
    % \right) \vert \Omega \rangle +\ldots \\
    % &= 
    \left({\rm A}^{\rm CCZ}_{\rm succ}\right)^2 \,\hat{a}_{w_{A}^1}^\dagger \,\hat{a}_{w_{A}^2}^\dagger\,\hat{a}_{w_{A}^3}^\dagger\,\hat{a}_{w_{A}^4}^\dagger \!\!\!\sum_{j_0,\ldots,j_4=\{0,1\}} (-)^{j_0\,j_1\,j_2+j_2\,j_3\,j_4}\,A_{j_0} B_{j_1} C_{j_2} D_{j_3} E_{j_4} \,\hat{a}_{w^0_{j_0}}^\dagger \hat{a}_{w^1_{j_1}}^\dagger\hat{a}_{w^2_{j_2}}^\dagger\hat{a}_{w^3_{j_3}}^\dagger\hat{a}_{w^4_{j_4}}^\dagger \,\vert \Omega \rangle
        +\ldots\,,
\end{equation}
where the dots contain all the terms that do not preserve the qubit structure. Indeed, we achieve the following desired evolution
\begin{equation}
    \vert q_0 \rangle_{\rm in} \otimes \vert q_1 \rangle_{\rm in} \otimes \vert q_2 \rangle_{\rm in} \otimes \vert q_3 \rangle_{\rm in} \otimes \vert q_4 \rangle_{\rm in} 
    \to \sum_{j_0,\ldots,j_4=\{0,1\}}  (-)^{j_0\,j_1\,j_2+j_2\,j_3\,j_4}\,A_{j_0} B_{j_1} C_{j_2} D_{j_3} E_{j_4}\,\vert j_0 \rangle \otimes \vert j_1 \rangle \otimes \vert j_2 \rangle \otimes \vert j_3 \rangle \otimes \vert j_4 \rangle \,,
\end{equation}
where $\vert q_0 \rangle_{\rm in} = \left( A_{0} \,\vert \mathsf{0} \rangle + A_{1} \,\vert \mathsf{1} \rangle \right)$, $\vert q_1 \rangle_{\rm in} = \left( B_{0} \,\vert \mathsf{0} \rangle + B_{1} \,\vert \mathsf{1} \rangle \right)$, $\vert q_2 \rangle_{\rm in} = \left( C_{0} \,\vert \mathsf{0} \rangle + C_{1} \,\vert \mathsf{1} \rangle \right)$, $\vert q_3 \rangle_{\rm in} = \left( D_{0} \,\vert \mathsf{0} \rangle + D_{1} \,\vert \mathsf{1} \rangle \right)$ and $\vert q_4 \rangle_{\rm in} = \left( E_{0} \,\vert \mathsf{0} \rangle + E_{1} \,\vert \mathsf{1} \rangle \right)$.

\end{appendices}

\begin{backmatter}

\bmsection{Funding}
Horizon 2020 Framework Programme (899368);
Provincia Autonoma di Trento through the Q@TN joint laboratory.

\bmsection{Acknowlegments}
The work was supported by the Horizon 2020 Framework Programme (899368) and by the Provincia Autonoma di Trento through the Q@TN joint laboratory.

\bmsection{Disclosures}
The authors declare no conflicts of interest.

\bmsection{Data availability}
No data were generated or analyzed in the presented research.

\end{backmatter}

\bibliography{CZandCCZ}

\end{document}